\documentclass[iop]{emulateapj}
\slugcomment{Accepted by ApJ, April 12, 2017}
\usepackage{amsmath,times,hyperref}

\shorttitle{Protostellar Evolution in Orion}
\shortauthors{Fischer et al.}

\begin{document}

\newcommand{\Spitzer}{\textit{Spitzer}}
\newcommand{\Herschel}{\textit{Herschel}}

\title{The \textit{Herschel} Orion Protostar Survey: Luminosity and Envelope Evolution}
\author{
William J. Fischer$^{1,2}$,
S. Thomas Megeath$^3$,
Elise Furlan$^4$,
Babar Ali$^5$,
Amelia M. Stutz$^{6,7}$,
John J. Tobin$^{8,9}$,\\
Mayra Osorio$^{10}$,
Thomas Stanke$^{11}$,
P. Manoj$^{12}$,
Charles A. Poteet$^1$,
Joseph J. Booker$^3$,
Lee Hartmann$^{13}$,\\
Thomas L. Wilson$^{14}$,
Philip C. Myers$^{15}$,
and Dan M. Watson$^{16}$
}
\affil{
$^1$Space Telescope Science Institute, Baltimore, MD, USA; wfischer@stsci.edu \\
$^2$NASA Goddard Space Flight Center, Greenbelt, MD, USA \\
$^3$Ritter Astrophysical Research Center, Department of Physics and Astronomy, University of Toledo, Toledo, OH, USA \\
$^4$IPAC, California Institute of Technology, Pasadena, CA, USA \\
$^5$Space Science Institute, Boulder, CO, USA \\
$^6$Max-Planck-Institut f\"ur Astronomie, Heidelberg, Germany \\
$^7$Departmento de Astronom\'ia, Facultad Ciencias F\'isicas y Matem\'aticas, Universidad de Concepci\'on, Concepci\'on, Chile \\
$^8$Leiden Observatory, Leiden University, Leiden, The Netherlands \\
$^9$Homer L. Dodge Department of Physics and Astronomy, University of Oklahoma, Norman, OK, USA \\
$^{10}$Instituto de Astrof\'isica de Andaluc\'ia, CSIC, Granada, Spain \\
$^{11}$European Southern Observatory, Garching bei M\"unchen, Germany \\
$^{12}$Department of Astronomy and Astrophysics, Tata Institute of Fundamental Research, Mumbai, India \\
$^{13}$Department of Astronomy, University of Michigan, Ann Arbor, MI, USA \\
$^{14}$National Science Foundation, Arlington, VA, USA \\
$^{15}$Harvard-Smithsonian Center for Astrophysics, Cambridge, MA, USA \\
$^{16}$Department of Physics and Astronomy, University of Rochester, Rochester, NY, USA
}

\begin{abstract}
The \Herschel\ Orion Protostar Survey obtained well-sampled 1.2 -- 870 \micron\ spectral energy distributions (SEDs) of over 300 protostars in the Orion molecular clouds, home to most of the young stellar objects (YSOs) in the nearest 500 pc.  We plot the bolometric luminosities and temperatures for 330 Orion YSOs, 315 of which have bolometric temperatures characteristic of protostars.  The histogram of bolometric temperature is roughly flat; 29\% of the protostars are in Class~0.  The median luminosity decreases by a factor of four with increasing bolometric temperature; consequently, the Class~0 protostars are systematically brighter than the Class~I protostars, with a median luminosity of 2.3~$L_\sun$ as opposed to 0.87~$L_\sun$.  At a given bolometric temperature, the scatter in luminosities is three orders of magnitude.  Using fits to the SEDs, we analyze how the luminosities corrected for inclination and foreground reddening relate to the mass in the inner 2500 AU of the best-fit model envelopes.  The histogram of envelope mass is roughly flat, while the median corrected luminosity peaks at 15~$L_\sun$ for young envelopes and falls to 1.7~$L_\sun$ for late-stage protostars with remnant envelopes.  The spread in luminosity at each envelope mass is three orders of magnitude.  Envelope masses that decline exponentially with time explain the flat mass histogram and the decrease in luminosity, while the formation of a range of stellar masses explains the dispersion in luminosity.
\end{abstract}


\section{INTRODUCTION}\label{s.intro}

\setcounter{footnote}{16}

For roughly the first 500,000 years in the formation of a young star \citep{eva09,dun14}, a rotating, infalling envelope feeds a circumstellar disk, which in turn accretes onto a hydrostatically supported central object.  Young stellar objects (YSOs) with such envelopes are known as protostars.  With observations over the last decade by the \Spitzer\ \citep{wer04} and \Herschel\ \citep{pil10} space telescopes, more than 1000 protostars and more than 4000 young stars that have lost their envelopes but retain disks have been identified in the nearest 0.5 kpc \citep{reb10,dun15,meg16}.  The Orion molecular clouds are home to 504 \Spitzer-identified candidate protostars \citep{meg16} and 16 additional \Herschel-identified candidates \citep{stu13,tob15}, easily making it the largest single collection of protostars in this volume.  

In the \Herschel\ Orion Protostar Survey (HOPS), a key program of the \Herschel\ Space Observatory, we obtained infrared (IR) imaging and photometry of over 300 of the Orion protostars at 70 and 160 \micron\ with the Photoconductor Array Camera and Spectrometer (PACS) instrument \citep{pog10} aboard \Herschel.  We supplemented our \Herschel\ observations with archival and newly obtained imaging, photometry, and spectra from 1.2 to 870~\micron, allowing modeling of the protostellar spectral energy distributions (SEDs) and images.  Details of the \Herschel\ photometry are presented in B.\ Ali et al.\ (in preparation), while the 1.2 to 870 \micron\ SEDs of the protostars are presented in \citet{fur16}.

With a sample of hundreds of protostars observed over three orders of magnitude in wavelength, we are able to reliably measure the bolometric properties of each source, constrain their underlying physical properties via modeling \citep{fur16}, and perform a statistical study of the evolution of protostellar envelopes.  Since the SEDs are strongly modified by the absorption and reprocessing of radiation from the central stars by dusty disks and infalling envelopes, the shape of an SED is expected to evolve as the protostar evolves (e.g., \citealt{ada87}).  To capture this evolution, YSOs were initially divided into classes based on the slopes $\alpha$ of their near-to-mid-IR SEDs from roughly 2 to 20 \micron\ \citep{lad87,gre94}, where $\alpha=\left(d\log\lambda S_\lambda\right)/\left(d\log\lambda\right)$, $\lambda$ is the wavelength, and $S_\lambda$ is the flux density at $\lambda$.  Class~I sources have $\alpha\ge0.3$, flat-spectrum sources have $-0.3\le\alpha<0.3$, Class~II sources have $-1.6\le\alpha<-0.3$, and Class~III sources have $\alpha<-1.6$.

The discovery of Class~0 objects \citep{and93}, which were difficult to detect in the mid-IR until the launch of \Spitzer, motivated additional criteria not based on the slope of the SED.  The bolometric temperature $T_{\rm bol}$, the effective temperature of a blackbody with the same mean frequency as the protostellar SED \citep{mye93}, was adopted to distinguish between Class~0 and Class~I sources.  Class~0 objects have $T_{\rm bol}<70~{\rm K}$, Class~I objects have $70~{\rm K}<T_{\rm bol}<650~{\rm K}$, and Class~II objects have $650~{\rm K}<T_{\rm bol}<2800~{\rm K}$ \citep{che95}. Flat-spectrum sources in the $\alpha$-based system are not explicitly included in this scheme, although \citet{eva09} suggest a range of 350 to 950 K, straddling the Class I/II boundary.

To classify the HOPS sample, \citet{fur16} adopted a hybrid approach, using $T_{\rm bol}$ to distinguish Class~0 objects from more evolved sources and using $\alpha$ (measured between 4.5 and 24 \micron) to classify these more evolved sources as Class~I, flat-spectrum, or Class~II objects.  They consider Class~0, I, and flat-spectrum objects to be protostars, while Class~II objects are post-protostellar, when the envelope has dissipated and only a circumstellar disk remains.  (See Section 7.2.3 of \citealt{fur16} for a small number of exceptions to this distinction.)  While \citet{hei15} found that only half of their flat-spectrum sources, which were selected based on the extinction-corrected 2 to 24 \micron\ spectral index, have envelopes detected in HCO$^+$, \citet{fur16} found that nearly all of the HOPS flat-spectrum sources have SEDs best fit with envelopes that are generally less massive than those of Class 0 and Class I protostars.

With model fits, \citet{fur16} found a systematic decrease in envelope density from Class~0 to Class~I to flat-spectrum protostars, with an overall decrease of a factor of 50.  This decrease is consistent with the interpretation that SED classes describe an evolutionary progression driven by the gradual dissipation of the envelope.  The classification, however, is affected by additional factors. Inclination can affect the SED, where a Class~I protostar viewed through an edge-on disk can have a lower $T_{\rm bol}$ than a Class~0 protostar viewed from an intermediate inclination angle.  Foreground reddening is a further complication, in that a more evolved object that lies behind extensive foreground dust may appear to have a more massive envelope (and therefore lower $T_{\rm bol}$) than it really does.  

To disentangle observational degeneracies in probing the evolution of envelopes, radiative transfer models have been employed to constrain physical parameters such as envelope density and mass.  Based on fits of models to SEDs, \citet{rob06} proposed the use of stages, where the stage refers to the underlying physical state probed by observations.  For protostars, Stage~0 refers to the period when the envelope mass $M_{\rm env}$ still exceeds the mass of the central object $M_*$, and Stage I refers to the period when $0<M_{\rm env}<M_*$.  The physical stages correspond only roughly to the observational classes \citep{dun14}.  Fitting models to the HOPS SEDs, \citet{fur16} tabulated the properties of the best-fit models.  They also analyzed uncertainties in the model fits, showing that, although models provide good fits to the data, the solutions are not necessarily unique, and degeneracies in model fit parameters can lead to large uncertainties.  For this reason, the use of model fits provides an alternative means of examining the evolution of protostars, but it does not fully replace the use of observational criteria such as SED class.

The bolometric luminosity and temperature (BLT) plot is a common evolutionary diagram for protostars first presented by \citet{mye93}, analogous to the Hertzsprung-Russell diagram for stars.  Data from the \Spitzer\ program ``From Molecular Cores to Planet-Forming Disks'' (c2d) were used to derive the BLT diagram for 1024 YSOs in five molecular clouds that are closer than Orion \citep{eva09}.  With the relative numbers of YSOs in each class, the c2d team estimated median lifetimes of 0.16 Myr for Class~0, 0.38 Myr for Class~I, and 0.40 Myr for the flat-spectrum phase, with small revisions downward after correcting for interstellar extinction.  The luminosities at each bolometric temperature were found to be spread over several orders of magnitude.

\citet{eva09} compared these findings to the models of \citet{you05}, which feature a constant envelope infall rate and are an extension of the \citet{shu77} inside-out collapse model.  These models predict a small range of luminosities due to the formation of a range of stellar masses, and these luminosities are large compared to those typically observed.  For Class I protostars, the model luminosities are of order 10 $L_\sun$, while the observed ones are generally $<3$~$L_\sun$. This is consistent with the classic luminosity problem first noted with \textit{Infrared Astronomical Satellite} data by \citet{ken90}.

\citet{dun10} explored the ability of various modifications to the \citet{you05} model to reproduce the broad luminosity spread in the c2d BLT diagram.  As suggested in the paper that originally established the luminosity problem, \citet{dun10} found that the most successful modification was to add episodic accretion, where the infalling matter from the envelope accumulates in the disk.  The growing mass in the disk contributes little to the observed luminosity until it abruptly accretes onto the star, yielding a luminosity outburst.  Explanations for this phenomenon typically invoke disk instabilities, either thermal instabilities (e.g., \citealt{bel94}), the magnetorotational and gravitational instabilities acting in concert \citep{zhu09,zhu10}, or the accretion of clumps formed when the accumulation of envelope material causes the disk to fragment \citep{vor05,vor15}.  The luminosity in this scenario is thus usually smaller than predicted by \citet{you05}, but it agrees when averaged over both the quiescent and outburst modes.

\citet{off11} compared the broad spread in protostellar luminosities, also noted for Orion and other clouds in the nearest 1 kpc by \citet{kry12}, to the predictions of various star-formation models.  They found that models with a roughly constant accretion time, not a constant accretion rate, better reproduced the observed luminosity distributions. They also found that tapered models, where the mass infall rate diminishes at late times, were able to produce a distribution where the typical Class 0 luminosity is equal to or greater than the typical Class I luminosity.  These contrasting approaches to resolving the luminosity problem, episodic (stochastic) accretion on one hand and slow (secular) variations of the accretion rate on the other, were discussed in detail by \citet{dun14} and are difficult to disentangle observationally.

Here we present the BLT diagram of the Orion protostars, showing the distribution for the largest number to date of completely sampled SEDs at a common distance.  We then use the radiative transfer modeling by \citet{fur16} to plot the inner envelope masses of the protostars, investigate luminosity evolution across the protostellar phase, and interpret these findings with simple models of star formation.  Section~2 describes the sample selection and observations, Section~3 presents BLT diagrams for the entire HOPS sample as well as for regions within Orion, Section~4 introduces model-based diagnostics that trace luminosity and envelope evolution, Section~5 interprets the evolutionary diagrams, and Section~6 contains our conclusions.

\section{SAMPLE DEFINITION AND OBSERVATIONS}

For our analysis we adopt the same sample of 330 YSOs as \citet{fur16}, who have tabulated their coordinates, photometry, properties, and model fits.  These are candidate protostars that were targeted by our \Herschel\ observations and detected in the PACS 70 \micron\ images.  They are spread over the Orion A and B molecular clouds from declinations of $-8^\circ50'$ to $1^\circ54'$ and from right ascensions of $5^{\rm h}33^{\rm m}$ to $5^{\rm h}55^{\rm m}$.  The Orion Nebula itself is excluded due to saturation in the \Spitzer\ maps used for sample selection.

We used photometry and spectra from several archival and new surveys to construct the SEDs of sources in the sample, which are plotted in \citet{fur16}.  Near-IR photometry from the Two Micron All Sky Survey (2MASS; \citealt{skr06}) and mid-IR photometry from \Spitzer\ appear in \citet{meg12}.  Mid-IR spectra from the \Spitzer\ Infrared Spectrograph (IRS) are plotted in \citet{fur16}.  The \Herschel\ photometry, including 70 and 160~\micron\ photometry from HOPS and 100 \micron\ photometry from the public archive, and photometry at 350 and 870 \micron\ from the Atacama Pathfinder Experiment (APEX) appear in \citet{fur16}.  The \Herschel\ and APEX surveys will be discussed in greater detail by B.\ Ali et al.\ (in preparation) and T.\ Stanke et al.\ (in preparation), respectively.

Using $T_{\rm bol}$, the 4.5 to 24 \micron\ spectral slope, and qualitative assessment of the SEDs, \citet{fur16} found 92 Class~0 protostars, 125 Class~I protostars, 102 flat-spectrum protostars, and 11 Class~II objects among the 330 sources.  In the fitting of their SEDs, six of the 330 sources were found to lack envelope emission.

\begin{deluxetable*}{lcccccc}
\tablecaption{Target Classification\label{t.target}}
\tablewidth{\hsize}
\tablehead{\colhead{} & \colhead{Dec Range} & \colhead{Sources in} & \colhead{Number of} & \colhead{Class~0} & \colhead{Class~I} & \colhead{Fraction of Protostars} \\ 
\colhead{Region} & \colhead{(J2000; $^\circ$)} & \colhead{Sample}& \colhead{Protostars} & \colhead{Protostars} & \colhead{Protostars} & \colhead{in Class~0\tablenotemark{1}}}
\startdata
All & $(-8.9,+1.9)$ & 330 & 315 &  91 & 224 & $0.29\pm0.03$ \\
L 1641 & $(-8.9,-6.1)$ & 173 & 160 &  32 & 128 & $0.20\pm0.03$ \\
ONC & $(-6.1,-4.6)$ &  79 &  77 &  27 &  50 & $0.35\pm0.05$ \\
Orion B & $(-2.5,+1.9)$ &  78 &  78 &  32 &  46 & $0.41\pm0.06$
\enddata
\tablenotetext{1}{Uncertainties are those in the quantity $n_0 / ( n_0 + n_{\rm I} )$, where the Class 0 and Class I counts are $n_0$ and $n_{\rm I}$ and are assumed to have uncertainties $\sqrt{n_0}$ and $\sqrt{n_{\rm I}}$.}
\end{deluxetable*}

\section{BOLOMETRIC LUMINOSITIES AND TEMPERATURES}

With far-IR photometry, we sample the peaks of the protostellar SEDs and thus derive more accurate bolometric properties than otherwise possible.  In a BLT diagram, the bolometric luminosity $L_{\rm bol}$ is the luminosity integrated over the observed SED. It can differ from the true luminosity of the protostar due to inclination along the line of sight, where a protostar viewed through its edge-on disk will appear less luminous than the same protostar viewed along its axis of rotation, or due to extinction.  The bolometric temperature is \begin{equation}T_{\rm bol} = 1.25\times10^{-11}\int_0^\infty \nu S_\nu\, d\nu \bigg/ \int_0^\infty S_\nu\, d\nu~{\rm K~Hz^{-1}},\end{equation} where $\nu$ is the frequency and $S_\nu$ is the flux density at that frequency \citep{mye93}. It is as low as 20 K for the most embedded protostars \citep{stu13} and increases as the envelope and disk accrete onto the star, reaching the effective temperature of the central star when circumstellar material is negligible.  For a given protostar, $T_{\rm bol}$ also depends on the inclination.

We obtained bolometric luminosities and temperatures by trapezoidal integration under the available photometry and IRS spectra using \texttt{tsum.pro} from the IDL Astronomy Users' Library.\footnote{See \url{http://idlastro.gsfc.nasa.gov/}.}  Upper limits are ignored, and the IRS spectra are rebinned to 16 fluxes.  For the luminosities, we assume a distance of 420 pc to Orion based on high-precision parallax measurements of non-thermal sources in the Orion Nebula region \citep{san07,men07,kim08,kou17}.

\subsection{The BLT Diagram}

\begin{figure}
\includegraphics[width=\hsize]{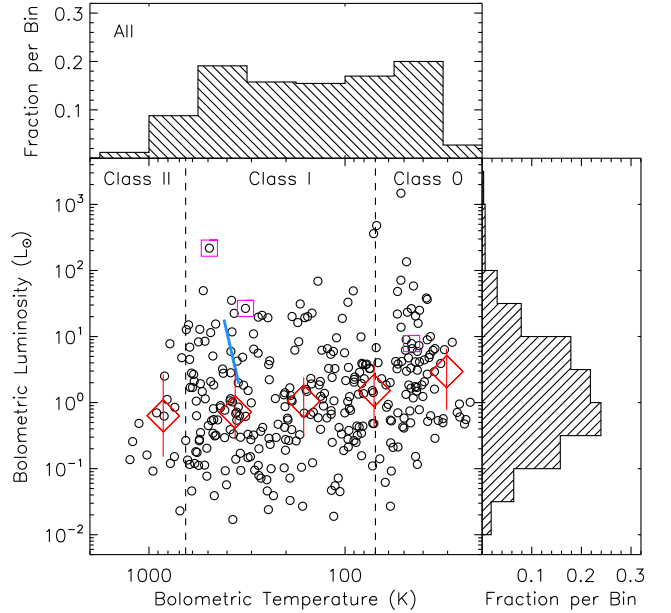}
\caption{Bolometric luminosities and temperatures of all 330 YSOs in the sample.  Dashed lines show the traditional divisions into Class~0, Class~I, and Class~II.  Large diamonds show the median luminosities in each of five bins that are equally spaced in $\log\ T_{\rm bol}$, and the solid vertical lines show the interquartile luminosity ranges.  The histograms show the marginal distributions for luminosity and temperature.  The blue line connects the pre- and post-outburst positions of HOPS 223; the symbol that happens to lie near its midpoint represents a different protostar.  Pink boxes mark the post-outburst locations of the other three luminosity outbursts in the sample:\ HOPS~376 is the more luminous of the two Class~I outbursts, HOPS 388 is the other, and HOPS 383 is the Class~0 outburst.\label{f.blt}}
\end{figure}

The BLT diagram for the 330 HOPS targets treated in this paper appears in Figure~\ref{f.blt}, and classification statistics appear in Table~\ref{t.target}.  There are 91 Class~0 sources, 224 Class~I sources, and 15 Class~II sources.  Of the 315 protostars (Class~0 and Class~I objects), 29\% are in Class~0. Because we consider only $T_{\rm bol}$, these counts differ slightly from the results of \citet{fur16}, reviewed in Section 2.  

While the standard classification scheme by $T_{\rm bol}$ does not contain a flat-spectrum category, the sources classified as such by \citet{fur16} have $T_{\rm bol}$ ranging from 83 to 1200~K with the middle 80\% falling between 190 and 640 K; the mean is 431 K.  This distribution features lower temperatures than that of \citet{eva09}, who found that the middle 79\% of their flat sources have $T_{\rm bol}$ between 350 and 950 K with a mean of 649 K.  After correcting for extinction, the middle 77\% of their flat sources have $T_{\rm bol}^\prime$ between 500 and 1450~K with a mean of 844 K.  Compared to the results from \citet{fur16}, their larger temperatures before extinction correction are likely due to different definitions of the class, where \citeauthor{fur16} use the spectral index between 4.5 and 24 \micron\ and \citet{eva09} use the index between 2 and 24 \micron.  The first definition allows sources that have rising SEDs from 2 to 4.5 \micron\ (a sign of extinction, either intrinsic to the source or foreground) and thus have relatively lower $T_{\rm bol}$ to be classified as flat.  In Section~\ref{s.diagnostics} we show how $T_{\rm bol}$ is dependent on foreground reddening, particularly for sources with low envelope densities.  Differences among authors in the definition of spectral slope and the means of correction for foreground reddening, if any, add uncertainty in the claimed range of $T_{\rm bol}$ for flat-spectrum sources.

In Figure~\ref{f.blt} we also display the histogram of $L_{\rm bol}$, which is the protostellar luminosity function of the sample, and the histogram of $T_{\rm bol}$.  As seen in Table~\ref{t.lum}, the bolometric luminosities of the HOPS protostars range over nearly five orders of magnitude, from 0.017 to 1500~$L_\sun$, with a mean of 13~$L_\sun$ and a median of 1.1~$L_\sun$.  The luminosity shows a clear peak near 1~$L_\sun$, a width at half-maximum in $\log (L/L_\sun)$ of 2, and a tail extending beyond 100~$L_\sun$.  The overall shape is similar to that determined by the extrapolation of \Spitzer\ photometry by \citet{kry12} for the Orion molecular clouds as well as for other giant molecular clouds forming massive stars, such as Cep OB3 and Mon R2.  The protostellar luminosity function derived from the \Spitzer\ c2d and Gould Belt surveys by \citet{dun13} peaks at a higher luminosity. That diagram uses luminosities corrected for extinction, but it shows a similar width to the Orion luminosity function.  

In contrast, the histogram of $T_{\rm bol}$ is quite flat.  Each of the bins between 30 and 600~K contains 15 to 20\% of the sample.  Note that the drop-off for Class~II sources is a selection effect due to the focus of HOPS on protostars.  Across Orion, the number of Class~II sources exceeds the number of protostars by a factor of three \citep{meg16}.

To examine how luminosity depends on evolutionary state, we divide the sample into five bins of equal spacing in $\log\ T_{\rm bol}$.  Table~\ref{t.bltreg} shows the five bins, the number of sources in each bin, and the median and interquartile range of their luminosities.  (The interquartile range is the difference between the third and first quartiles of the distribution.)  These results also appear as the large red diamonds in Figure~\ref{f.blt}, with the interquartile ranges plotted as vertical red bars.  They show a monotonic decrease in the median $L_{\rm bol}$ with increasing $T_{\rm bol}$ across the full range of protostars.  They also show a wide range of luminosities in each bin, a spread of three orders of magnitude.  The monotonic decline in median luminosities and broad spread in luminosities are the two most salient properties of the HOPS BLT diagram.

\begin{figure}
\includegraphics[width=\hsize]{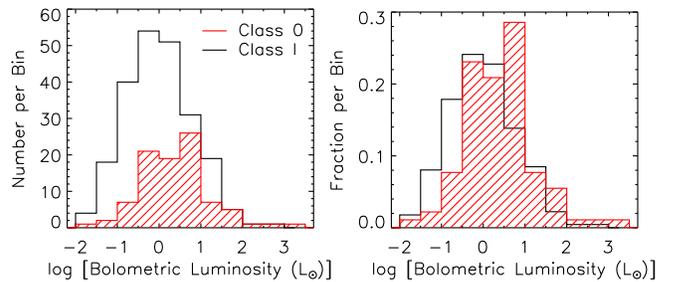}
\caption{Histograms of bolometric luminosity for the 91 Class~0 and 224 Class~I protostars. The left panel shows the number per bin, and the right panel shows the fraction of each class per bin to facilitate comparison.\label{f.histo}}
\end{figure}

This decrease in luminosity can also be shown by dividing the sample into Class~0 and Class~I protostars. The Class~0 luminosities are larger, ranging from 0.027 to 1500~$L_\sun$ with a mean of 30~$L_\sun$ and a median of 2.3~$L_\sun$.  The Class~I luminosities range from 0.017 to 360~$L_\sun$ with a mean of 6.5~$L_\sun$ and a median of 0.87~$L_\sun$.  A two-sample Kolmogorov-Smirnov (KS) test reveals a probability of only $5.5\times10^{-4}$ that the Class~0 and Class~I luminosity histograms were drawn from the same distribution.  Figure~\ref{f.histo} shows the histograms of the two classes, plotted both as the number per bin and as the fraction of each class per bin.  As we discuss in Section~\ref{s.bias}, the difference in luminosity is unlikely to be due to the effects of incompleteness and extinction on the BLT diagram.

\begin{deluxetable}{lcccc}
\tablecaption{Bolometric Luminosity Statistics for Protostars\label{t.lum}}
\tablewidth{\hsize}
\tablehead{\colhead{} & \colhead{Minimum} & \colhead{Maximum} & \colhead{Median} & \colhead{Mean} \\ 
\colhead{Region} & \colhead{($L_\sun$)} & \colhead{($L_\sun$)} & \colhead{($L_\sun$)} & \colhead{($L_\sun$)}}
\startdata
All     & 0.017 & 1500 & 1.1  & 13   \\
L 1641  & 0.017 &  220 & 0.70 &  5.0 \\
ONC     & 0.046 &  360 & 2.4  & 12   \\
Orion B & 0.027 & 1500 & 1.5  & 30   \\
\cutinhead{Class~0 Only}
All     & 0.027 & 1500 & 2.3  & 30   \\
L 1641  & 0.027 &  140 & 1.6  & 12   \\
ONC     & 0.25  &   38 & 4.2  &  9.1 \\
Orion B & 0.062 & 1500 & 2.9  & 65   \\
\cutinhead{Class~I Only}
All     & 0.017 &  360 & 0.87 &  6.5 \\
L 1641  & 0.017 &  220 & 0.69 &  3.7 \\
ONC     & 0.046 &  360 & 1.9  & 14   \\
Orion B & 0.027 &   33 & 0.89 &  5.7
\enddata
\end{deluxetable}

\subsection{Dependence of the BLT Diagram on Region}

\begin{figure}
\includegraphics[width=\hsize]{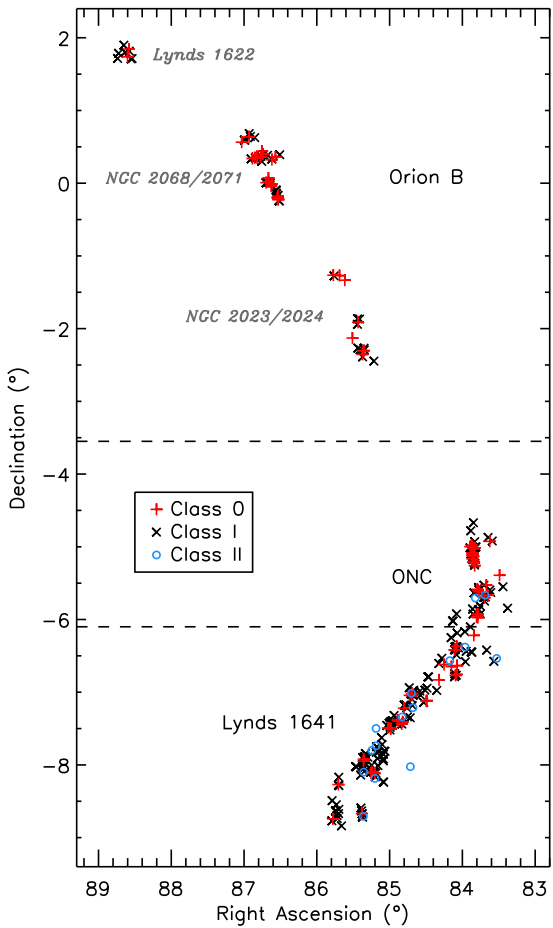}
\caption{Locations of the 330 sources within Orion and the dividing lines that separate them into regions.  Sources are coded by $T_{\rm bol}$ class as shown.  Names of the regions used for statistics are printed in black, while names of the Orion B subregions are printed in gray italics.\label{f.orion}}
\end{figure}

With 330 sources, we can divide Orion into regions and retain enough protostars in each to examine BLT trends as a function of location or environment.  Due to the roughly north-south alignment of the Orion molecular clouds, we define the regions simply as declination ranges. Figure~\ref{f.orion} shows how the 330 sources, color-coded by $T_{\rm bol}$ class, are divided into regions.  This division into groups is beneficial, because we can compare BLT diagrams for two separate molecular clouds within the Orion OB association:\ Orion A and B.

\begin{deluxetable*}{lcccccccc}
\tablecaption{Median Bolometric Luminosities by Region and Bolometric Temperature\tablenotemark{1}\label{t.bltreg}}
\tablewidth{\hsize}
\tablehead{\colhead{Range of} & \multicolumn{2}{c}{All} & \multicolumn{2}{c}{L 1641} & \multicolumn{2}{c}{ONC} & \multicolumn{2}{c}{Orion B} \\ \colhead{$T_{\rm bol}$ (K)} & \colhead{Number} & \colhead{$\left<L_{\rm bol}\right>$ ($L_\sun$)} & \colhead{Number} & \colhead{$\left<L_{\rm bol}\right>$ ($L_\sun$)} & \colhead{Number} & \colhead{$\left<L_{\rm bol}\right>$ ($L_\sun$)} & \colhead{Number} & \colhead{$\left<L_{\rm bol}\right>$ ($L_\sun$)}}
\startdata
(20,   46) & 49 & 2.9 (5.8) & 13 & 1.2 (3.4) & 17 & 6.6 (7.2) & 19 & 2.1 (5.3) \\
(46,  110) & 90 & 1.5 (3.4) & 44 & 0.94 (2.0) & 20 & 3.3 (4.4) & 26 & 2.2 (4.3) \\
(110,  240) & 67 & 1.1 (2.1) & 42 & 0.71 (1.1) & 14 & 1.9 (5.4) & 11 & 0.89 (19.) \\
(240,  550) & 86 & 0.72 (3.0) & 47 & 0.63 (1.7) & 20 & 1.1 (4.8) & 19 & 0.54 (6.4) \\
(550, 1300) & 38 & 0.63 (2.5) & 27 & 0.62 (1.5) &  8 & 2.1 (4.4) &  3 & 2.8 (15.)
\enddata
\tablenotetext{1}{Luminosities in parentheses are the interquartile range in each bin.}
\end{deluxetable*}

The HOPS sources north of $-2.5^\circ$ are part of the Orion B molecular cloud \citep[e.g.,][]{wil05,bal09}.  This consists of three distinct fields:\ the Lynds 1622 field, the field containing the NGC 2068/2071 nebulae, and the field containing the NGC 2024/2023 nebulae \citep{meg12}.  These fields contain two clusters, a number of groups, and relatively isolated stars \citep{meg16}.  While there is some disagreement as to whether these are parts of a single coherent cloud, they have similar distance and velocity, so we combine all 78 sources in Orion B for this work. 

Orion A contains HOPS sources south of $-4.6^\circ$.  (Due to the gap between Orion A and B, there are no HOPS sources between $-2.5^\circ$ and $-4.6^\circ$.)  We divide Orion A into two regions, setting the boundary at $-6.1^\circ$.  The northern region is the Orion Nebula Cluster (ONC).   While the Orion Nebula itself contains no HOPS sources due to saturation in the 24 \micron\ \Spitzer\ band used to identify them, the outer regions of the ONC are rich in HOPS protostars.  It contains 79 sources.  Our ONC field, while larger than some definitions of the ONC and encompassing Orion Molecular Cloud (OMC) 2, 3, and 4, approximates the boundaries in \citet{car00} and \citet{meg16}. The southern region of Orion A is Lynds 1641 (L~1641); it contains 173 sources, including multiple clusters, groups, and isolated protostars.  Dividing the Orion A cloud thus gives us the opportunity to compare the BLT diagram of a rich cluster to that of a cloud dominated by smaller groups, clusters, and relatively isolated stars.

\begin{figure}
\includegraphics[width=\hsize]{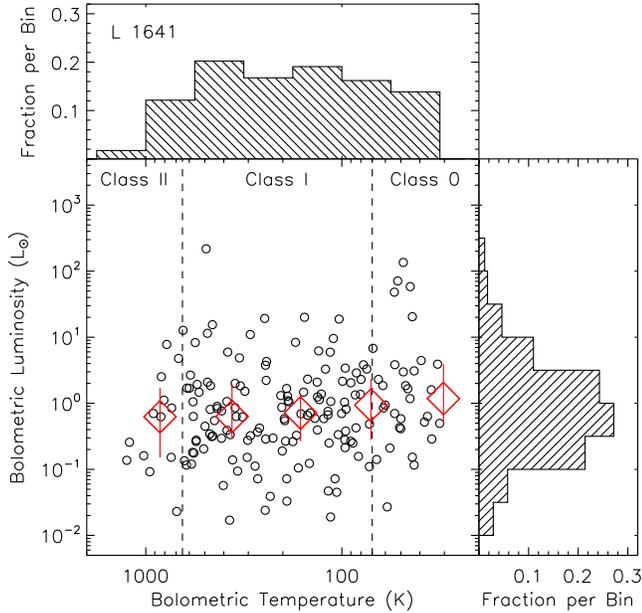}
\caption{Bolometric luminosities and temperatures of the 173 sources in L 1641 (between declinations $-8.9^\circ$ and $-6.1^\circ$). Temperature bins are the same as in Figure~\ref{f.blt}.\label{f.blt_l1641}}
\end{figure}

\begin{figure}
\includegraphics[width=\hsize]{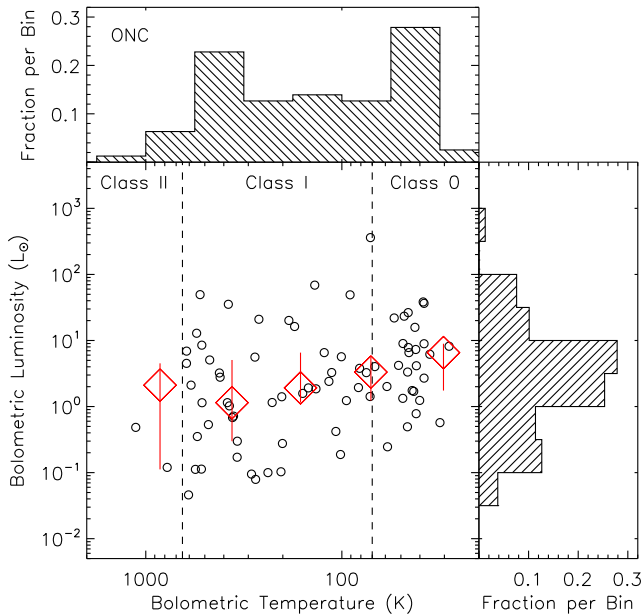}
\caption{Bolometric luminosities and temperatures of the 79 sources in the ONC (between declinations $-6.1^\circ$ and $-4.6^\circ$). Temperature bins are the same as in Figure~\ref{f.blt}.\label{f.blt_onc}}
\end{figure}

\begin{figure}
\includegraphics[width=\hsize]{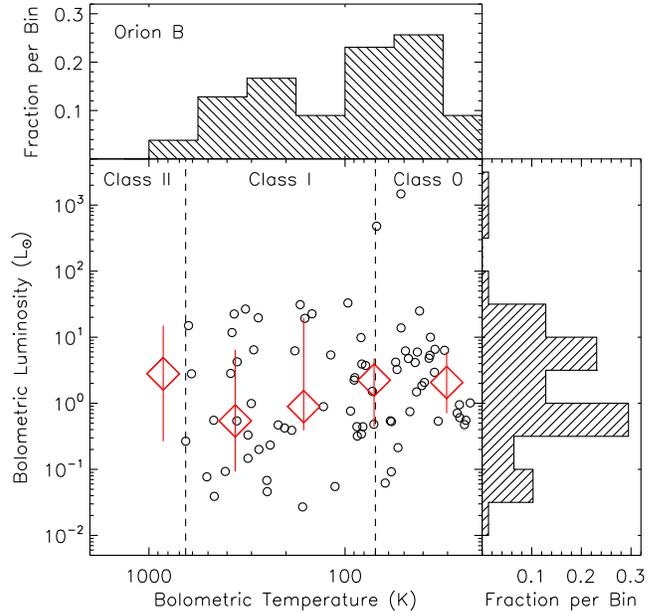}
\caption{Bolometric luminosities and temperatures of the 78 sources in Orion B (between declinations $-2.5^\circ$ and $1.9^\circ$). Temperature bins are the same as in Figure~\ref{f.blt}.\label{f.blt_orionb}}
\end{figure}

Table~\ref{t.target} lists the regions and the number of sources, number of protostars of each class, and fraction of Class~0 protostars for each.  Tables~\ref{t.lum} and \ref{t.bltreg} give the luminosity statistics for each region, and Figures~\ref{f.blt_l1641} through \ref{f.blt_orionb} show the BLT diagrams for each region.  

The division of protostars between Class~0 and Class~I is similar among the three regions and the whole sample, but there are important differences.  The fraction of protostars in Class~0 increases from south to north, going from $0.20\pm0.03$ in L~1641 to $0.35\pm0.05$ in the ONC to $0.41\pm0.06$ in Orion B.  \citet{stu15} found a similar increase from south to north within L~1641 and the ONC.  The larger Class~0 fraction in Orion B meshes with the finding of \citet{stu13} that the fraction of sources that are PACS Bright Red Souces (PBRS, a class of extremely young protostars) is higher in Orion B (0.17) than in Orion A (0.01).

The typical bolometric luminosities of the protostars are largest in the ONC and smallest in L~1641, with the median luminosity declining from 2.4~$L_\sun$ in the ONC to 1.5~$L_\sun$ in Orion B to 0.70~$L_\sun$ in L~1641.  In each region, the median Class~0 source is more luminous than the median Class~I source by a factor ranging from 2.2 in the ONC to 3.3 in Orion~B.  In Orion B and L~1641, the {\em mean} bolometric luminosity is also larger in Class~0 than in Class~I.  This is not the case in the ONC; there, the high mean luminosity for Class~I protostars is mainly due to HOPS 370 (OMC 2 FIR 3; \citealt{mez90}; \citealt{ada12}).  It has $L_{\rm bol}=361~L_\sun$ and $T_{\rm bol}=71.5$~K, near the Class~0/I boundary.  Without this source, the mean luminosity for Class~I ONC sources is 7.3~$L_\sun$, less than that of the Class~0 protostars in the region.

We also show the luminosities in the five $T_{\rm bol}$ bins discussed above.  In each region, the bolometric luminosity decreases with increasing bolometric temperature, except for the bins of highest $T_{\rm bol}$ in the ONC and in Orion B, which contain very few sources, and between the two bins of lowest $T_{\rm bol}$ in Orion B.  The interquartile ranges vary between 1 and 8 $L_\sun$ for most bins, although the lightly populated bins for $T_{\rm bol} > 110$~K in Orion B have ranges up to 19 $L_\sun$.

\subsection{Luminosity Outbursts}

Five Orion protostars have been identified as outbursting sources.  (See \citealt{aud14} for a recent review of the outburst phenomenon in YSOs.) Reipurth 50 \citep{str93} lacks a HOPS identifier; it was saturated in the 4.5~\micron\ \Spitzer\ band used to find protostars when establishing the HOPS target catalog and is not part of the \citet{fur16} sample.  V883 Ori (HOPS 376; \citealt{str93}) and V1647 Ori (McNeil's Nebula; HOPS 388; \citealt{mcn04}) began their outbursts before they were imaged with \Spitzer.  The pre-outburst SED of HOPS 383 \citep{saf15} was faint and poorly sampled, and a firm estimate of its pre-outburst bolometric properties is impossible.  For HOPS 376, 383, and 388, the \citet{fur16} properties used here are based on only their post-outburst SEDs. They are shown with pink boxes in Figure~\ref{f.blt}.

The fifth outburst, V2775 Ori (HOPS 223; \citealt{car11,fis12}), has a well-sampled SED both before and after its outburst. \citet{fur16} tabulated its BLT properties based on its combined pre- and post-outburst SEDs, acknowledging that this gives unreliable numbers but aiming for a uniform treatment of the large sample.  We find a pre-outburst bolometric luminosity and temperature of 1.93~$L_\sun$ and 348~K, and we find post-outburst BLT properties of 18.0~$L_\sun$ and 414~K.  (Pre-outburst data are from Table 1 of \citealt{fis12}, while post-outburst data combine photometry from Table 2 of that paper with photometry derived from the 2011 IRTF spectrum presented therein.)  While the pre-outburst properties are less reliable due to a lack of photometry beyond 70 \micron, HOPS 223 is a member of Class~I at both epochs.  The pre- and post-outburst positions of HOPS 223 in BLT space are connected with a blue line in Figure~\ref{f.blt}. They are not used in the calculations of statistics; for this we retain the bolometric properties tabulated by \citet{fur16}, which place the object in the cluster of three points near 20~$L_\sun$ and 250~K.

\subsection{Effect of Incompleteness and Extinction\label{s.bias}}

When comparing the luminosities of the Class~I and Class~0 protostars, potential biases due to incompleteness and extinction must be considered. In the \Spitzer\ data, detection schemes can miss very deeply embedded Class~0 protostars with weak fluxes at wavelengths $\le24$ \micron.  To mitigate this source of incompleteness, \citet{stu13} augmented the HOPS sample with 70 \micron\ images acquired by \Herschel/PACS to find new protostars not identified with \Spitzer. They found that the original \Spitzer-based detection \citep{meg12,meg16} was not significantly incomplete, as there were only 15 likely protostars detected at 70~\micron\ that were missed in the \Spitzer\ sample of more than 300. \citet{tob15} subsequently found one more. The majority of the newly detected protostars (14/16) are located in L 1641 or Orion B, not in regions of high nebulosity like the ONC, indicating that these sources were not previously detected due to their unusually faint 24~\micron\ fluxes and not due to incompleteness in the \Spitzer\ data resulting from confusion with nebulosity.

Another concern is that the far-IR nebulosity may hinder the detection of faint protostars in the 70 \micron\ band.  However, the decrease in luminosity between the Class~0 and Class~I sources persists across various regions within Orion, including the high-background ONC and the low-background L~1641 (Figures \ref{f.blt_l1641} through \ref{f.blt_orionb}).  This suggests that the difference in the luminosities is not the result of incompleteness to faint Class~0 protostars.

A final potential bias in the data is that foreground extinction may lead to the misclassification of protostars.  \citet{stu15} studied the effects of extinction-driven misclassification of Class I and Class 0 protostars, both by foreground material and ``self-extinction'' due to inclination.  They find that when far-IR data are included in the SED analysis, as is the case here, the extinction-driven misclassification probability is negligible over statistical sample sizes such as ours.  Specifically, they find that for protostars with measured $T_{\rm bol}=70$~K (that is, borderline Class~0 YSOs), the probability of misclassification is $< 15\%$ with foreground extinction levels of $A_V=30$ mag and steeply decreases with lower extinction levels.  Furthermore, they find median extinction levels for HOPS protostars of $A_V=23.3$ mag in the ONC and $\sim 12$~mag in L~1641, indicating that misclassification of this type is not a concern when far-IR data are included in protostellar SED analysis.

A related concern is the potential misclassification of reddened Class II objects as flat-spectrum or Class I protostars.  \citet{fur16} classified the 330 sources with $T_{\rm bol}$, the 4.5 to 24 \micron\ slope, and qualitative assessment of the SEDs, finding that 319 are Class 0, I, or flat-spectrum.  Since $A_{[4.5]}$ is about 0.5 $A_{K_s}$ \citep{fla07}, the slope from 4.5 to 24 \micron\ is less influenced by foreground reddening than slopes that include data from shorter wavelengths.   Additionally, far-IR photometry exists for the entire sample, and far-IR emission is affected very little by extinction. If an envelope exists, the far-IR emission will be stronger than if there is just a disk, so envelope- and disk-dominated sources are more easily distinguishable with such data.  When modeling the sources, \citet{fur16} found that for 324 of the 330, the far-IR emission is best fit with a model that includes an envelope.  Our assessment, using only $T_{\rm bol}$, finds 315 Class 0, I, or flat-spectrum sources.  (Five sources that are Class I by $T_{\rm bol}$ alone are Class II in the multi-pronged analysis by \citeauthor{fur16}, and nine sources that are Class II by $T_{\rm bol}$ alone are Class~I or flat-spectrum in their analysis.)  While there is minor disagreement between an analysis limited to $T_{\rm bol}$ and one that uses additional information, multiple lines of evidence suggest that nearly all of our sources have protostellar envelopes.

\section{UNDERSTANDING PROTOSTELLAR EVOLUTION VIA SED MODELING}\label{s.model}

Modeling of the 330 SEDs is described in detail by \citet{fur16}.  Here we review the most important points.  The HOPS team created a grid of 3040 SED models, each viewed from ten inclinations, with the code of \citet{whi03}.  This code performs Monte Carlo simulations of radiative transfer through a dusty circumstellar environment. It uses an axisymmetric geometry and includes a central luminosity source, a flared disk with power-law scale height and radial density profiles, an envelope defined by the rotating spherical collapse model of \citet{ulr76}, and a bipolar envelope cavity with walls described by a polynomial expression.

The models sample parameters of interest in the study of protostars: 19 mass infall rates that scale the envelope density profile (including the case of no envelope), four disk radii, and five cavity opening (half-)angles.  The system luminosity can take on values between 0.05 and 600 $L_\sun$.  Other parameters, including the dust properties, are held constant, as described in \citet{fur16}.  The quality of the model fits is evaluated with the parameter $R$. This is a measure of the average, weighted, logarithmic deviation between the observed and model SEDs; the model with the minimum value of $R$ is the best-fit model.  \citeauthor{fur16}\ found that most protostars are well fit by models from the grid, although there are some degeneracies among model parameters, and the quality of the best-fit model for each protostar depends in part on how well-constrained the SED is.  They estimate the reliability of each model fit by examining the modes of parameter values of models within a certain range of the best-fit $R$.  We refer the reader to that paper for plots showing the quality of the fit to each object.

Among other results, \citet{fur16} report the modeled envelope mass inside 2500~AU for each source, which is a function of other model parameters as shown below.  In this section we show the utility of this mass in diagnosing envelope evolution.  We then show how differences between the total luminosities of protostars and their observed luminosities may be accounted for via SED modeling.  Finally, we examine the relationship between the evolutionary states and total luminosities of the HOPS sources using results from the fitting.

\subsection{Model-based Masses as an Envelope Diagnostic\label{s.diagnostics}}

Since a primary goal of studies of protostellar evolution is to track the flow of mass from the molecular cloud onto the central forming star, the envelope mass $M_{\rm env}$ remaining inside some radius $r$ is a useful diagnostic of envelope evolution.  The youngest protostars have massive envelopes, while Class II objects have little to no remnant envelope.  Further, the envelope mass is an easily understood quantity that changes in a straightforward way with the inclusion of outflow cavities and is independent of inclination angle.  While we expect the envelope mass within 2500 AU to be correlated with both the ultimate main-sequence mass of the  star and the age of the protostar, the envelope masses we model extend over four orders of magnitude, and the stars formed will mostly have masses that extend over about two orders of magnitude. Thus the envelope mass is mainly sensitive to age and is expected to be the intrinsic property that best traces age.

We set $r$, the radius inside which we consider the envelope mass, equal to 2500~AU. This corresponds to the 6\arcsec\ half-width at half maximum of the 160 \micron\ PACS beam at the distance of Orion.  This is the largest spatial scale probed by the HOPS point-source photometry near the expected peaks of the SEDs in the sample.  The analysis in Section 5.2 assumes that envelope material inside 2500~AU is participating in free-fall toward the star, which is expected to be the case for all but the youngest sources.  

The models we use assume axisymmetry, with deviations from spherical symmetry due to rotational flattening of the envelope and the presence of outflow cavities.  These are characterized, respectively, by the centrifugal radius $R_C$ and the cavity opening angle $\theta_{\rm cav}$.  The centrifugal radius gives the outer radius at which the infalling envelope material accumulates onto the central Keplerian disk.  It may initially be equal to the outer radius of the disk, and this is assumed to be the case in our grid of models, although viscous spreading will cause the disk to expand outward.  The cavity opening angle is the angle from the pole to the cavity edge at a height above the disk plane equal to the envelope radius.  (See Figure~6 of \citealt{fur16} for a schematic illustration.)

\begin{figure}
\includegraphics[width=\hsize]{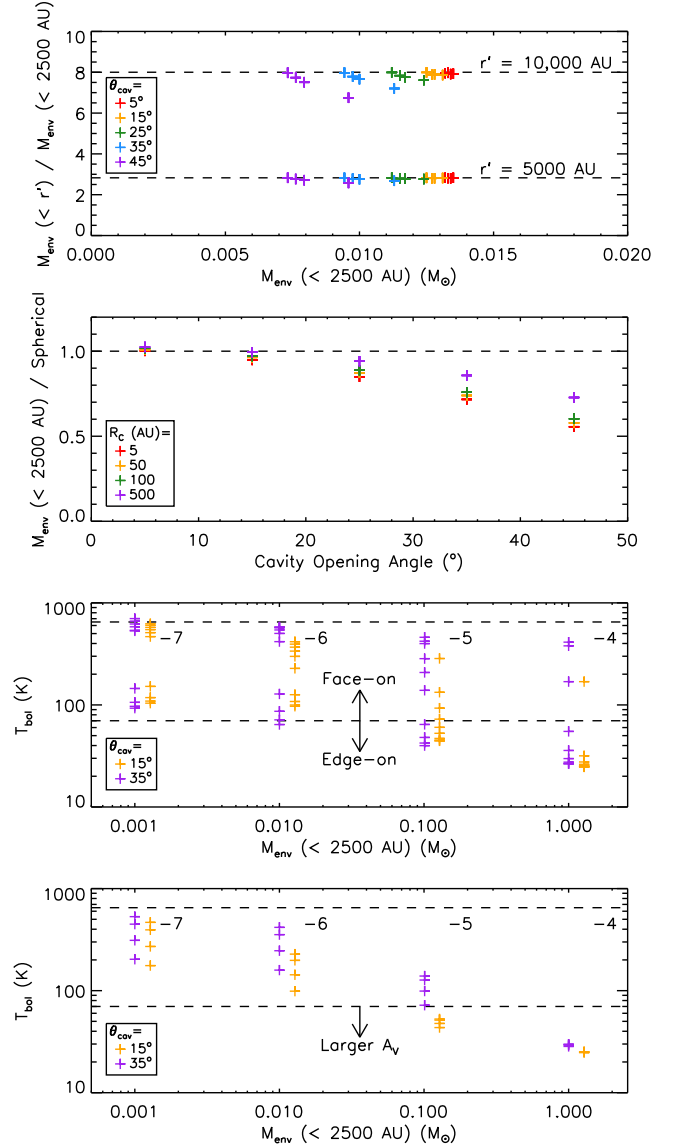}
\caption{{\em Top panel:} Ratio of envelope mass inside 5000 or 10,000 AU to that at 2500 AU, plotted against envelope mass inside 2500 AU for models with $\dot{M}_{\rm env}=10^{-6}~M_\sun~{\rm yr}^{-1}$.  Points of the same color but increasing mass correspond to increasing $R_C$.  Points of differing color correspond to different $\theta_{\rm cav}$. The envelope mass is not dependent on inclination angle. The dashed lines show the ratios expected for a strict $r^{1.5}$ mass dependence.  {\em Second panel:} Comparison of envelope masses from the grid to results for the angle-averaged solution with no cavity. The grid masses depend mildly on $R_C$ and dramatically on cavity angle.  {\em Third panel:} Bolometric temperature versus envelope mass inside 2500 AU for a selection of models with the indicated $\log \dot{M}_{\rm env}$ (in $M_\sun$ yr$^{-1}$) and cavity opening angles.  The spread in $T_{\rm bol}$ at each envelope mass is due to varying inclination angle, from low $T_{\rm bol}$ near edge-on to high $T_{\rm bol}$ near face-on.  Dashed lines mark the traditional boundaries between SED classes.  {\em Bottom panel:} Same as above, except the results are shown only for an inclination angle of $63^\circ$ as the SED is subjected to foreground extinction ranging from $A_V=0$ to 19.0 mag.\label{f.massenv}}
\end{figure}

The masses inside 2500 AU are easily scaled to other radii $r^\prime$, as seen in the top panel of Figure~\ref{f.massenv}.  To a close approximation, the masses can be multiplied by $(r^\prime/2500~{\rm AU})^{1.5}$.  Points of the same color and increasing mass show the effect of increasing $R_C$ from 5 to 500 AU.  Points of different colors show the effect of changing $\theta_{\rm cav}$ from 5$^\circ$ to 45$^\circ$.  The largest discrepancies between the actual masses within 5000 or 10,000 AU and those extrapolated from 2500~AU occur for large $R_C$ and $\theta_{\rm cav}$. 

In the case of spherical symmetry, we can relate the mass to the infall rate, which is often used to parameterize models \citep{whi03}.  The relationship is 
\begin{equation}
\begin{split}
M_{\rm env}\left(<r\right)=0.105\:M_\sun\bigg(\frac{\dot{M}_{\rm env}}{10^{-6}\:M_\sun\,{\rm yr}^{-1}}\bigg)\!\bigg(\frac{M_*}{0.5\:M_\sun}\bigg)^{-1/2}\\\times\bigg(\frac{r}{10^4\:{\rm AU}}\bigg)^{3/2}\label{e.mdot},
\end{split}
\end{equation}
where $\dot{M}_{\rm env}$ is the rate at which matter from the envelope accumulates onto the disk and $M_*$ is the mass of the central star.  Note that this assumes a constant, spherical infall, with the dominant mass being the central protostar.

Another common model parameter is $\rho_1$, the envelope density at 1 AU in the limit of no rotation \citep{ken93}. The relationship between envelope mass and $\rho_1$ is
\begin{equation}
M_{\rm env}\left(<r\right)=0.139~M_\sun \bigg(\frac{\rho_1}{10^{-14}~{\rm g~cm}^{-3}}\bigg)\bigg(\frac{r}{10^4~{\rm AU}}\bigg)^{3/2}.
\end{equation}
Like the envelope infall rate, this quantity does not account for changes in the cavity opening angle.  For envelopes with $R_C\gg1~{\rm AU}$, it also gives densities much larger than actually exist in the envelope \citep{fur16}. 

We explore the effect of deviations from spherical symmetry on envelope mass in the second panel of Figure~\ref{f.massenv}, which shows the effect of centrifugal radius $R_C$ and cavity opening angle $\theta_{\rm cav}$ on the mass inside 2500 AU.  For a rotating envelope ($R_C>0$), the mass depends weakly on $R_C$ for $R_C\ll r$, inducing a small vertical spread in points of different colors.  The mass depends more strongly on the cavity opening angle $\theta_{\rm cav}$, since large fractions of the envelope are removed with increasing $\theta_{\rm cav}$.  The mass is reduced by up to 45\% for the largest cavity opening angle.  The top two panels of Figure~\ref{f.massenv} show results for the models in our SED grid with $\dot{M}_{\rm env}=10^{-6}~M_\sun~{\rm yr}^{-1}$ and with $M_*=0.5~M_\sun$, but the behavior is the same for other $\dot{M}_{\rm env}$ and $M_*$.  This demonstrates the value of an envelope diagnostic that includes the effects of different cavity opening angles; reporting only the envelope infall rate or a representative density can be misleading.

In the rest of Figure~\ref{f.massenv}, we show how these masses compare to $T_{\rm bol}$.  While it has the advantage of being directly measurable from observed SEDs, $T_{\rm bol}$ depends strongly on source inclination and foreground reddening.  In the third panel of Figure~\ref{f.massenv}, we compare $T_{\rm bol}$ to  $M_{\rm env}\left(<2500~{\rm AU}\right)$ for selected models with $\dot{M}_{\rm env}$ of $10^{-7}$, $10^{-6}$, $10^{-5}$, and $10^{-4}~M_\sun~{\rm yr}^{-1}$ and cavity opening angles of $15^\circ$ and $35^\circ$.  For all models, there is a large spread in $T_{\rm bol}$ as the inclination runs from $18^\circ$ (highest $T_{\rm bol}$) to $87^\circ$ (lowest $T_{\rm bol}$), in some cases crossing the traditional boundaries between SED classes.  

In the bottom panel of Figure~\ref{f.massenv}, we compare $T_{\rm bol}$ to $M_{\rm env}\left(<2500~{\rm AU}\right)$ for the same models, except the inclination angle is held constant at $63^\circ$ and the foreground reddening is varied.  The largest $T_{\rm bol}$ in each case is for $A_V=0$ mag, and the bolometric temperature decreases as $A_V$ increases.  We show results for $A_V$ at the first, second, and third quartiles of the distribution used to model the HOPS sources, or $A_V=2.5$, 9.0, and 19.0 mag.  Varying $A_V$ over this range has less of an effect than varying the inclination angle over its entire range, but the spread is still several hundred K for the least massive envelopes.

\begin{figure}
\includegraphics[width=\hsize]{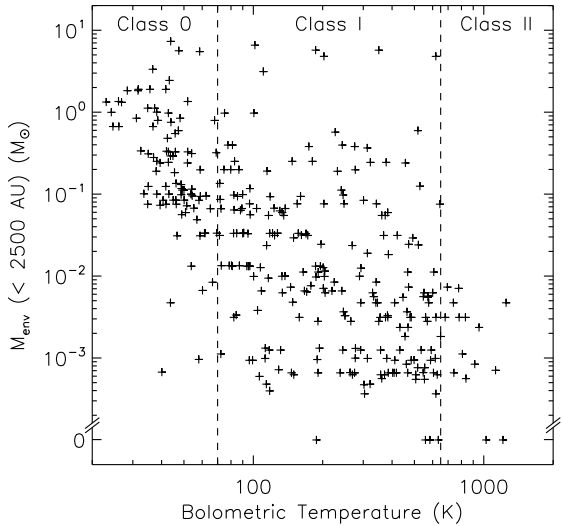}
\caption{Model-derived envelope mass within 2500 AU versus observed bolometric temperature for the 330 YSOs.  The dashed lines mark the divisions into SED classes.\label{f.menvtbol}}
\end{figure}

The envelope mass inside a particular radius is dependent on the assumed density distribution within the disk and envelope.  The SED models presented by \citet{fur16} fit the observations well, suggesting that the density distributions are plausible if not necessarily unique.  Compared to $T_{\rm bol}$, this mass is an alternative diagnostic of envelope evolution that is insensitive to inclination angle and foreground reddening.  In Figure~\ref{f.menvtbol} we plot the envelope mass within 2500 AU for each best-fit model against the bolometric temperature of the observed SED.  There is a weak anticorrelation between the two, with substantial scatter due to the dependence of $T_{\rm bol}$ on not only envelope mass, but also on source inclination and foreground reddening.

\subsection{Model-based Total Luminosities}

\begin{figure}
\includegraphics[width=\hsize]{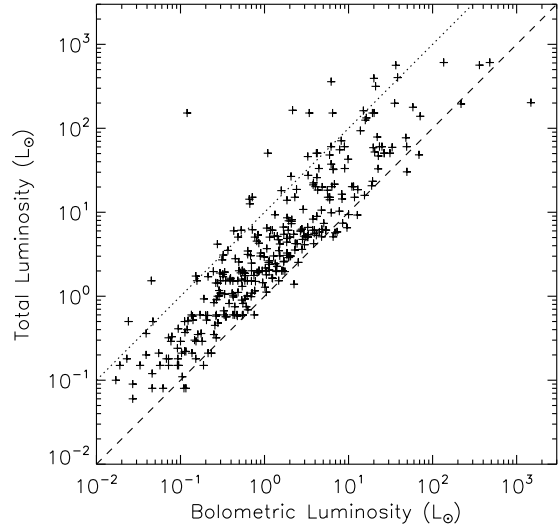}
\caption{Total (model-derived) luminosity versus bolometric (observed) luminosity for the 330 YSOs.  The dashed line marks equality, and the dotted line marks the case where the total luminosity is ten times the bolometric luminosity.  Cases with $L_{\rm tot} \gg L_{\rm bol}$ are addressed in Section 5.2.\label{f.ltotbol}}
\end{figure}

The total luminosity of a protostar generally differs from its bolometric luminosity due to foreground extinction and inclination.  Foreground extinction reduces the flux at all wavelengths (although trivially at far-IR wavelengths and longer), so correcting for this always increases the luminosity from its observed value.  Inclination can affect the luminosity in either direction.  Converting an observed flux to a luminosity involves multiplying by $4\pi$ sr, which assumes that the source is isotropic.  Due to high extinction by a circumstellar disk in the (approximate) equatorial plane of the star and low extinction through the cavity aligned with the rotation axis, protostars are brighter when viewed along their rotational axes.  Thus, multiplying by $4\pi$ sr overestimates the luminosity of a face-on protostar and underestimates the luminosity of an edge-on protostar.  (See Figure 7 of \citealt{fur16} for an example of how a protostellar SED changes with inclination angle.)

We correct for these effects with SED fitting.  In short, the colors of a protostar shortward of 70 \micron\ are sensitive to inclination, while the colors at longer wavelengths are sensitive to envelope mass, and fitting attempts to break the degeneracy between the two by simultaneously accounting for both wavelength regimes \citep{ali10}.  We subsequently analyze the total luminosity of the best-fit model from \citet{fur16} rather than the observed luminosity.

Figure~\ref{f.ltotbol} compares the modeled total luminosity~$L_{\rm tot}$ to the observed bolometric luminosity~$L_{\rm bol}$ for the 330 YSOs in the sample.  The quantity $\log (L_{\rm tot}/L_{\rm bol})$ has a mean of 0.47 and a standard deviation of 0.39, consistent with the ratios generally being greater than unity.  In almost all cases, the total luminosity is larger than the bolometric luminosity, because foreground extinction always reduces the bolometric luminosity, while inclination effects can either inflate or reduce the bolometric luminosity.  To consider the effect of inclination alone, we define $L_{\rm mod}$ as the integrated luminosity of the best-fit SED at the best-fit inclination, with the modeled foreground extinction removed.  The quantity $\log (L_{\rm tot}/L_{\rm mod})$ has a mean of 0.19 and a standard deviation of 0.36.  This shows that inclination tends to reduce the bolometric luminosity from the total luminosity; configurations that are sufficiently edge-on to reduce it are more likely than configurations that are sufficiently face-on to increase it.

\subsection{Total Luminosity versus Envelope Mass}

\begin{figure}
\includegraphics[width=\hsize]{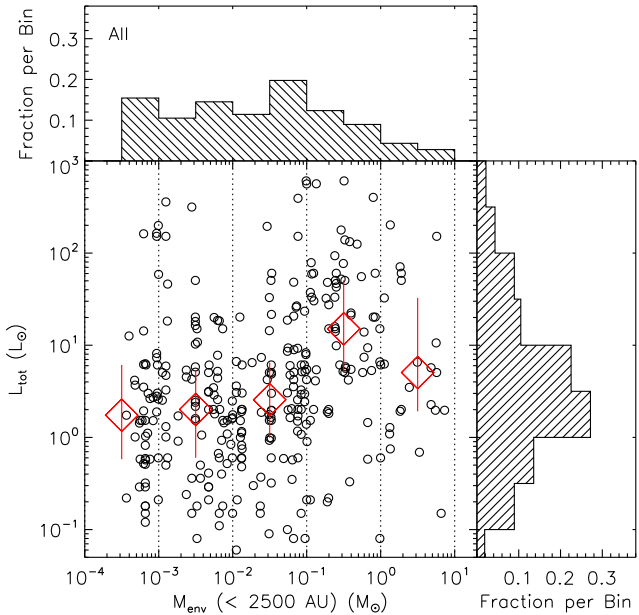}
\caption{Total luminosity versus envelope mass (TLM) inside 2500 AU for 324 protostars across the entire HOPS survey region.  (Six sources with $M_{\rm env} = 0$ in their best-fit models are excluded.)  The histograms show the marginal distributions for luminosity and mass.  Large diamonds show the median luminosities in each of the mass bins indicated by dotted vertical lines, and the solid vertical lines show the 
interquartile luminosity ranges.\label{f.tlm}}
\end{figure}

In Figure~\ref{f.tlm}, we plot the total luminosity~$L_{\rm tot}$ and envelope mass inside 2500 AU of the best-fit SED model assigned to each source, creating a total luminosity versus mass (TLM) diagram.  Models in the grid have luminosities of 0.1, 0.3, 1, 3, 10, 30, 100, or 300 $L_\sun$, and the luminosity is adjusted by a factor between 0.5 and 2 to improve the SED fit \citep{fur16}.  Therefore, the possible luminosities extend continuously from 0.05 to 600~$L_\sun$.  The mass inside 2500 AU is set by the envelope infall rate, the centrifugal radius, and the cavity opening angle.  The possible nonzero masses extend from $3.6\times10^{-4}$ to 10 $M_\sun$; there are 898 unique masses over this range.  The fractional change from one mass to the next largest ranges from 10$^{-4}$ to 0.36 with a median of $3\times10^{-4}$.

Of the 330 sources, six are fit with models that contain no envelope.  These are excluded from the analysis of how luminosity changes with envelope mass.  The 324 remaining sources have total luminosities ranging from 0.06 to 600~$L_\sun$, roughly the same as the allowed range, and they have envelope masses ranging from $3.6\times10^{-4}$ $M_\sun$ (the minimum nonzero mass possible) to 7.3 $M_\sun$.  The median total luminosity of all 324 sources is 3.0~$L_\sun$, larger than the median bolometric luminosity of 1.1~$L_\sun$, and the median envelope mass inside 2500~AU is 0.03~$M_\sun$.

We divide the sources into bins by envelope mass, with the edges of the bins at $10^{-4}$, $10^{-3}$, 0.01, 0.1, 1, and 10 $M_\sun$.  While there is substantial scatter in the luminosities at each envelope mass, the median luminosities in each bin show a clear trend.  They rise from 5.1~$L_\sun$ for the most massive envelopes to a peak of 15~$L_\sun$ in the bin extending from 0.1 to 1~$M_\sun$, and then they diminish from 2.6 to 1.7~$L_\sun$ over the three remaining bins.  The number of sources in each bin and their median luminosities and interquartile ranges are reported in Table~\ref{t.medlum}.

With medians and interquartile ranges, it can be difficult to assess whether the progression in luminosity with envelope mass is statistically significant.  Therefore we also ran two-sample KS tests on the luminosity distributions in each pair of mass bins to assess the likelihood that these luminosities were drawn from the same underlying distribution.  From least massive to most massive, the KS probabilities for the first and second, second and third, and first and third bins were, respectively, 92\%, 81\% and 39\%, consistent with our claim that there is little evolution in luminosity over the least massive envelopes. The probabilities that the fourth bin was drawn from the same distribution as any of the first three bins are all between 10$^{-7}$ and 10$^{-6}$, indicative of a statistically significant decline from the fourth to the third bin.  The KS probabilities for the fifth (most massive) bin compared to the first through fourth bins, are, respectively, 8\%, 20\%, 52\%, and 9\%.

\begin{figure}
\includegraphics[width=\hsize]{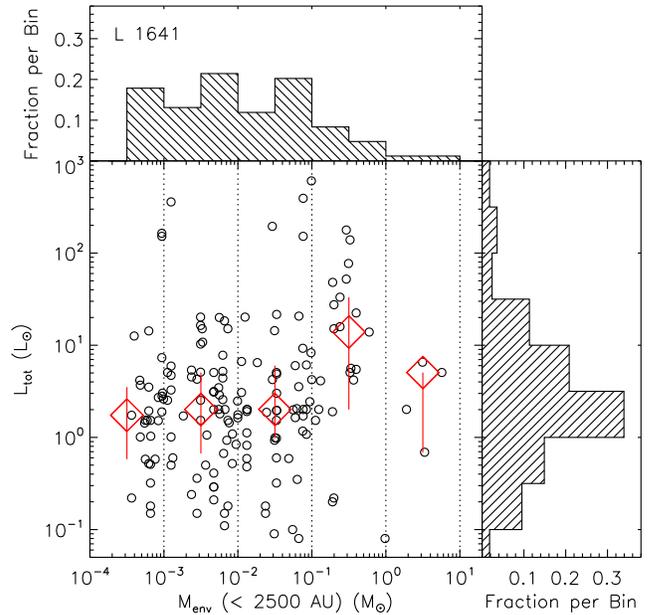}
\caption{Total luminosity versus envelope mass inside 2500 AU for 168 protostars in L 1641.\label{f.tlm_l1641}}
\end{figure}

\begin{figure}
\includegraphics[width=\hsize]{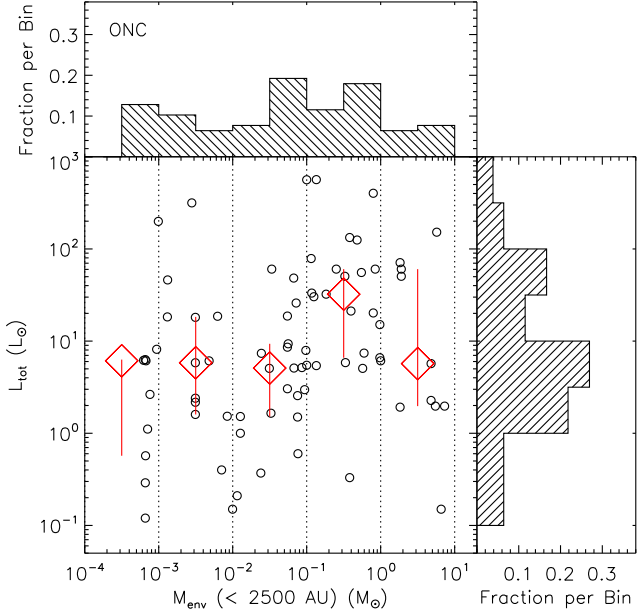}
\caption{Total luminosity versus envelope mass inside 2500 AU for 78 protostars in the ONC.\label{f.tlm_onc}}
\end{figure}

\begin{figure}
\includegraphics[width=\hsize]{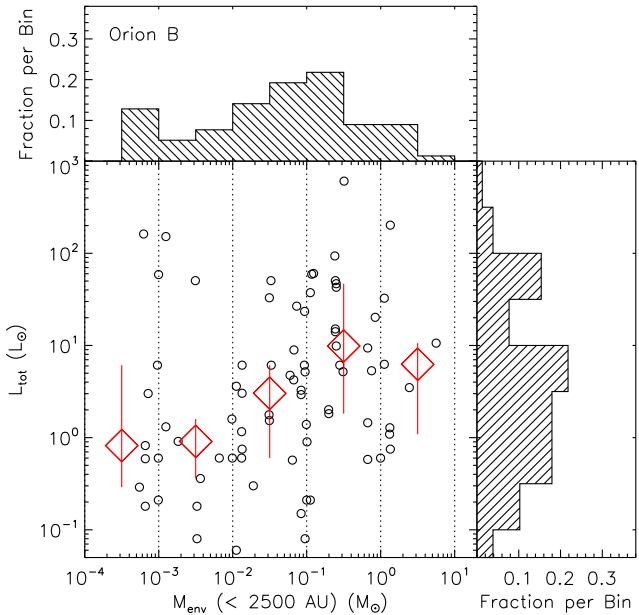}
\caption{Total luminosity versus envelope mass inside 2500 AU for 78 protostars in Orion B.\label{f.tlm_orionb}}
\end{figure}

Figures~\ref{f.tlm_l1641} through \ref{f.tlm_orionb} show TLM diagrams for each of our three defined regions, and Table~\ref{t.medlum} summarizes the median total luminosities as a function of envelope mass by region.  While the median total luminosity of all 324 sources with envelopes is 3.0~$L_\sun$, this quantity varies from region to region.  It is largest in the ONC, at 6.2~$L_\sun$, and it falls to 3.3~$L_\sun$ in Orion~B and 2.0~$L_\sun$ in L~1641.  In all three regions, the median luminosity is largest for envelopes between 0.1 and 1~$M_\sun$, again with much scatter.  The median luminosity falls for the next most massive bin and then tapers or remains roughly constant over the two least massive bins.

The total luminosity and envelope mass determine different properties of the SEDs.  The former determines the overall flux level, while the latter roughly sets the amount of emission at mid- to far-IR wavelengths.  To examine whether any degeneracies in the model fits may drive the reported trend of luminosity with envelope mass, for each source we consider the spread in total luminosities and inner envelope masses of all models that have $R<R_{\rm best}+2$, where the subscript refers to the best-fit model.  The number of fits that satisfy this criterion varies from source to source, but on average it allows the 448 best fits per source.  When considering all models that satisfy this criterion for all 330 sources, the standard deviation of $L/L_{\rm best}$ is 0.31 orders of magnitude, and the standard deviation of $M/M_{\rm best}$ is 1.29 orders of magnitude.  While large, these ratios are not correlated. Instead, they are slightly anti-correlated, with a correlation coefficient of $-0.23$. Therefore, uncertainty in the model fitting is unlikely to be the source of the trends discussed in this and subsequent sections.

\begin{deluxetable*}{lcccccccc}
\tablecaption{Median Total Luminosities by Region and Envelope Mass inside 2500 AU\tablenotemark{1}\label{t.medlum}}
\tablewidth{\hsize}
\tablehead{\colhead{Range of} & \multicolumn{2}{c}{All} & \multicolumn{2}{c}{L 1641} & \multicolumn{2}{c}{ONC} & \multicolumn{2}{c}{Orion B} \\ \colhead{$M_{\rm env}$ ($M_\sun$)} & \colhead{Number} & \colhead{$\left<L_{\rm tot}\right>$ ($L_\sun$)} & \colhead{Number} & \colhead{$\left<L_{\rm tot}\right>$ ($L_\sun$)} & \colhead{Number} & \colhead{$\left<L_{\rm tot}\right>$ ($L_\sun$)} & \colhead{Number} & \colhead{$\left<L_{\rm tot}\right>$ ($L_\sun$)}}
\startdata
($10^{0}$,$10^{1}$)   &  22 &  5.1 (31.) &   4 &  5.1 (4.4) &  10 &  5.7 (58.) &   8 &  6.3 (9.5) \\
($10^{-1}$,$10^{0}$)  &  67 &  15. (45.) &  22 &  14. (31.) &  22 &  32. (54.) &  23 &  9.9 (45.) \\
($10^{-2}$,$10^{-1}$) & 103 &  2.6 (5.5) &  53 &  2.0 (5.0) &  23 &  5.1 (7.8) &  27 &  3.0 (5.5) \\
($10^{-3}$,$10^{-2}$) &  82 &  2.0 (4.8) &  59 &  2.0 (4.4) &  13 &  5.8 (17.) &  10 &  0.91 (1.2) \\
($10^{-4}$,$10^{-3}$) &  50 &  1.7 (5.5) &  30 &  1.7 (2.9) &  10 &  6.1 (5.7) &  10 &  0.82 (5.8) \\
\tableline \rule{0pt}{2.5ex}
All                   & 324 &  3.0 (9.8) & 168 &  2.0 (4.7) &  78 &  6.2 (31.) &  78 &  3.3 (15.)
\enddata
\tablenotetext{1}{Luminosities in parentheses are the interquartile range in each bin.}
\end{deluxetable*}

\section{ANALYSIS OF LUMINOSITY AND ENVELOPE EVOLUTION}

In this section, we discuss three trends apparent in the BLT diagram that persist when we switch from observed parameters to intrinsic properties estimated via SED modeling.  These are \begin{itemize}\item the relatively flat distributions of bolometric temperature and envelope mass when considering the entire sample, \item the decrease in luminosity with decreasing envelope mass or increasing bolometric temperature, and \item the broad scatter in luminosity at each envelope mass or bolometric temperature.\end{itemize}

\subsection{The Flat Distribution of Envelope Mass}

The flat histogram of bolometric temperature noted in Section 3 persists when we transition to the envelope mass in Figure~\ref{f.tlm}.  The fraction of the objects in each bin varies between 10 and 20\% for masses between $3 \times 10^{-4}$ and 0.3 $M_\sun$.  At larger masses the histogram declines; there are fewer objects with envelopes of $\sim1$ $M_\sun$ or greater inside 2500~AU.  These presumably form higher-mass stars; our consideration of the initial mass function in Section 5.3 suggests that such massive stars and envelopes should be rare.

The relatively flat histogram suggests that $dN/d(\log M_{\rm env})$ is constant.  Expanding this expression, \begin{equation} \frac{dN}{d(\log M_{\rm env})} = \frac{dN}{dt} \frac{dt}{d(\log M_{\rm env})}\end{equation} is constant.  The first term on the right is the star-formation rate; if this is constant, then $d(\log M_{\rm env})/dt$ is also constant.  This implies an exponential decline in the envelope mass with time, which also suggests a roughly exponential decline in $\dot{M}$, the envelope infall rate.

This form for the infall rate is motivated by the work of \citet{bon96}, \citet{mye98}, \citet{sch04}, and \citet{vor10}. It is a consequence of the rate being roughly proportional to the remaining envelope mass, as expected if the rate equals the mass divided by some characteristic time, for example, the free-fall time for the mass within a given radius.  In contrast, \citet{oso99} found that the formation of the most massive stars (early B types and hotter) is best modeled with infall rates that increase with time.  Some of the most massive envelopes in the HOPS sample may exhibit this feature; however, our focus is on the lower-mass objects that dominate the sample.

\subsection{Decreasing Luminosities with Evolution\label{s.evo}}

\begin{figure*}
\includegraphics[width=\hsize]{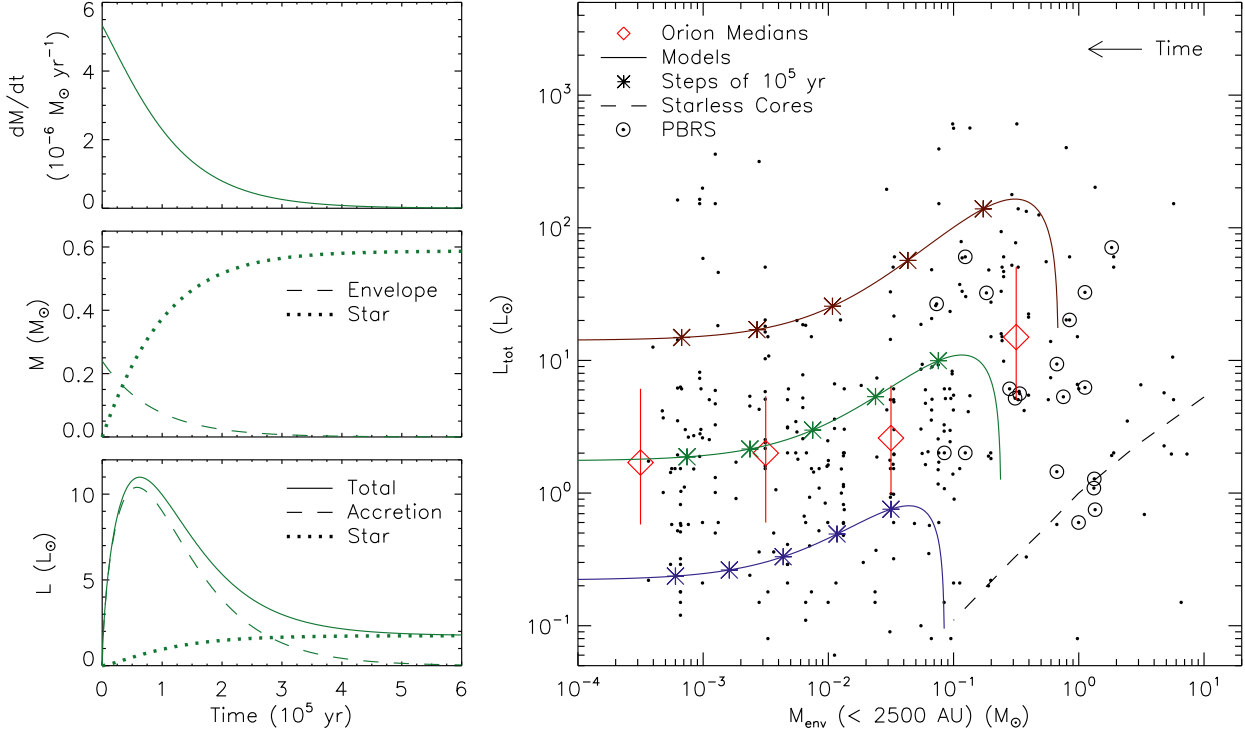}
\caption{{\em Left:} From top to bottom, the infall rate, masses (envelope and stellar), and luminosities (total, accretion, and stellar) for a model protostar.  {\em Right:} The TLM diagram from Figure~\ref{f.tlm} with model tracks overplotted.  HOPS sources that are PBRS are circled.  The median luminosity for the highest mass bin is not shown, since we are not comparing it to the model tracks.  The central curve corresponds to the model depicted in the left panels, while the upper and lower curves correspond to cases that form stars with different masses.  From bottom to top, the final stellar masses are 0.12, 0.58, and 2.8 $M_\sun$.  Time increases from right to left, with asterisks marking times from 10$^5$ yr to $5 \times 10^5$ yr in steps of 10$^5$ yr.  The dashed curve shows the luminosities for externally heated spherical starless cores ranging in mass from 0.1 to 10 $M_\sun$.\label{f.model}}
\end{figure*}

In the TLM diagram, the steep drop in median luminosity from the 0.1 -- 1 $M_\sun$ mass bin to the next less massive one, followed by a more gradual decline, suggests an exponentially declining form for the dependence of the protostellar luminosity on envelope mass.  This is also seen in the BLT diagram, where there is a slow decrease of $L_{\rm bol}$ with increasing $T_{\rm bol}$.  Here we show how this feature of the BLT and TLM diagrams is a consequence of the exponentially declining envelope masses hypothesized above.

For consistency with results from the radiative transfer model, we consider the envelope mass within a radius of 2500~AU (dropping the $<2500$ AU notation for simplicity).  The time dependence of the envelope mass is defined as \begin{equation}M_{\rm env}(t)=M_{\rm env,0}\times e^{-t\ln 2/t_H},\end{equation} where $t$ is the time elapsed, $M_{\rm env,0}$ is the initial mass, and $t_H$ is the time it takes for the mass to fall to half its initial value (the half-life).

The infall rate is $\dot{M}=M_{\rm env}/t_{\rm ff}$, where $t_{\rm ff}$ is the free-fall time within that radius, \begin{equation}t_{\rm ff} = \frac{\pi\,(2500\ {\rm AU})^{3/2}}{2\sqrt{2G(M_{\rm env}+M_*)}}.\end{equation}  This can be expressed as $t_{\rm ff}=q\,(M_{\rm env}+M_*)^{-1/2}$, where $q=2.2\times10^4$ yr $\sqrt{M_\sun}$. Then $\dot{M}=M_{\rm env}\,(M_{\rm env}+M_*)^{1/2}/q$.  At each time step, the stellar mass $M_*$ is $\int_0^t \dot{M}(t)\,dt$.

For comparison with the TLM diagram, we also calculate the luminosity as a function of time.  This is the sum of the stellar luminosity~$L_*$ and the accretion luminosity~$L_{\rm acc}$.  For the stellar luminosity we use a fit to the model tracks of \citet{sie00} for stars of mass 0.1 to 3 $M_\sun$ at a model age of $5\times10^5$ yr, \begin{equation}L_*=3.1\left(\frac{M_*}{0.9\,M_\sun}\right)^{1.34}.\label{e.lum}\end{equation} 

The accretion luminosity is $\eta\,GM_*\dot{M}/R_*$, where $G$ is the gravitational constant and $\eta$ is a factor of order unity that depends on the details of the accretion process and characterizes how much of the accretion energy is radiated away.  We do not explicitly include a circumstellar disk in this model, although the disk may act as a mass reservoir such that the accretion rate onto the star is not instantaneously equal to the envelope infall rate.  Here we set $\eta=0.8$, which is typical of accreting young stars \citep{mey97}.  The stellar radius is again a fit to the \citeauthor{sie00}\ models, \begin{equation}R_*=3.2\left(\frac{M_*}{0.9\,M_\sun}\right)^{0.34}.\end{equation}

The parameters of the central star are a source of uncertainty in this effort.  There are few observational constraints on stellar masses and radii for deeply embedded protostars, age is an ambiguous quantity at early times, and the accretion history of a given protostar is expected to have an important influence on its properties \citep{bar17}.  For simplicity, we adopt the \citet{sie00} models at a stated age of $5\times10^5$ yr.  These authors' models and the ``hybrid'' accretion case of \citeauthor{bar17}\ are similar at an age of 1 Myr, the earliest time at which the latter are tabulated.

In Figure~\ref{f.model}, we explore the relationship between total luminosity and envelope mass under these assumptions for three cases that produce stars of differing final masses $M_{*,f}$.  We arrange the simulations to yield final stellar masses such that the final stellar luminosities bracket the median total luminosity in the bin with the lowest envelope masses.  These luminosities are at the tenth, fiftieth, and ninetieth percentiles of the distribution, or 0.21, 1.7, and 14 $L_\sun$.  According to Equation (\ref{e.lum}), the final stellar masses are then 0.12, 0.58, and 2.8~$M_\sun$.  

The free parameters are the initial envelope mass inside 2500 AU and the half-life for the envelope mass.  These are chosen to yield the final stellar masses of interest and to reach the lowest envelope masses in the 0.5 Myr expected lifetime for protostars \citep{eva09}.  The inital masses are 0.085, 0.24, and 0.69 $M_\sun$.  (The initial envelope mass inside 2500~AU can be less than the final stellar mass due to the infall of material from beyond 2500 AU.)  The half-lives are 0.07, 0.06, and 0.05 Myr, respectively.  The characteristics of all three models are shown in Table~\ref{t.modelprop}.  The left panels of Figure~\ref{f.model} show the infall rate, stellar mass, envelope mass, stellar luminosity, accretion luminosity, and total luminosity for the case where $M_{\rm env,0}=0.24$~$M_\sun$ and $M_{*,f}=0.58$~$M_\sun$.  The right panel shows the path of this model through TLM space as well as those of the models with smaller and larger final stellar masses.

\begin{deluxetable}{lccc}
\tablecaption{Model Properties\label{t.modelprop}}
\tablewidth{\hsize}
\tablehead{\colhead{Property} & \colhead{Low} & \colhead{Medium} & \colhead{High}}
\startdata
$M_{\rm env,0}$ ($M_\sun$) & 0.085 & 0.24 & 0.69 \\
$\dot{M}_0$ (10$^{-6}$ $M_\sun$ yr$^{-1}$) & 1.1 & 5.3 & 26 \\
$t_H$ (Myr) & 0.07 & 0.06 & 0.05 \\
$M_{*,f}$ ($M_\sun$) & 0.12 & 0.58 & 2.8
\enddata
\end{deluxetable}

The luminosity is dominated by accretion over most of each track.  Since $R_*\propto M_*^{0.34}$, the accretion luminosity is proportional to $M_*^{0.66}\dot{M}$.  It initially rises quickly due to the increasing mass of the star and then falls off slowly due to the decline in the infall rate, leveling out at the luminosity of a star with the resulting final mass.  These curves are qualitatively similar to those shown in Figure 6b of \citet{and00}.

Uncertainties in the properties of the central star affect the plotted curves in a straightforward way.  For example, if the radius of the central star is actually 10\% larger than assumed, the total luminosity at early times will be 10\% smaller than assumed, when it is dominated by accretion and is inversely proportional to the stellar radius.  The total luminosity at late times will be 20\% larger than assumed, when it is dominated by the star and is approximately proportional to the stellar radius squared.  Both such effects are small compared to the range of the logarithmic luminosity axis in Figure~\ref{f.model}, and the qualitative shapes and positions of the curves will not change appreciably for discrepancies of this magnitude.

In the right panel of Figure~\ref{f.model}, none of the models pass through the median total luminosity in each mass bin, as one would expect if the median total luminosities represented a typical protostar as it moved through the various stages of envelope evolution.  By bracketing the majority of the data points, the models instead show how exponentially declining infall rates that produce a range of stellar masses can account for luminosities that, on average, decrease with evolution but are widely scattered.

The asterisks in Figure~\ref{f.model}, which show the positions of the models at 1, 2, 3, 4, and $5 \times 10^5$ yr, are equally spaced in $\log M_{\rm env}$ for each model.  Every 100,000 years, the envelope mass drops by 63\%, 69\%, or 75\% for the low-, medium-, and high-mass models, respectively.  This is consistent with the roughly flat distributions in $\log T_{\rm bol}$ (Figure~\ref{f.blt}) and $\log M_{\rm env}$ (Figure~\ref{f.tlm}).

Some outliers warrant additional attention.  First we look at those with high masses and low luminosities.  The dashed curve in the figure shows the predicted luminosities for spherical starless cores that range in mass from 0.1 to 10 $M_\sun$ and have uniform temperature 15~K, radius 2500 AU, and power-law density profile with exponent $-2$.  Objects much less luminous and more massive than this curve are inconsistent with centrally illuminated sources of the given mass.  Visual inspection of the \Herschel\ images indicates that in the extreme cases and many of those near the curve, the envelope masses are likely overestimated due to the inclusion of mass in the aperture that is not part of the protostellar envelope.  Another issue may be limitations to the grid of radiative transfer models and degeneracies that could yield unphysical parameters.  We do not attempt to explain these cases with our model, although we note that, without them, the median luminosity at earlier times is even higher, supporting the scenario of an early period of rapid accretion.  The circled protostars in Figure~\ref{f.model} are PBRS, the extremely young protostars discovered by \citet{stu13}.  For PBRS near the curve, visual inspection of the \Herschel\ and APEX images available in \citet{stu13} reveals bright point sources consistent with protostars that have large envelope masses and are truly at a young evolutionary stage.

Objects in the opposite corner of the TLM diagram, with total luminosities greater than 30~$L_\sun$ and envelopes less massive than 0.01 $M_\sun$, are also far from the regime covered by the model tracks.  Such a population of luminous, late-stage protostars does not appear in the observational BLT diagram (Figure~\ref{f.blt}).  These ten sources typically have model-derived luminosities much larger than their bolometric luminosities; eight of them have~$L_{\rm tot}/L_{\rm bol} > 10$.  These large ratios are due either to nearly edge-on orientations or to very large model-derived extinctions $A_V$ along the line of sight to the source.  In four cases, $A_V$ exceeds 50 mag.  \citet{fur16} judge the quality of the fit for each parameter by comparing the best-fit value to the mode of all fits within some range of acceptability.  For all of these sources, $A_V$ is not well constrained by this measure. 

The large extinctions may in some cases be due to the outer regions of a nearby protostellar envelope.  One of the sources in this group, HOPS 165, was modeled as such in \citet{fis10} due to its being only 5500 projected AU from HOPS 203.  Two others are also within 6000 projected AU of another HOPS source.

In their analysis of scattered-light images of the HOPS protostars, J. J. Booker et al.\ (in preparation) found that protostars that are undetected at 1.6 \micron, which usually correspond to Class 0 sources, have larger bolometric luminosities than protostars that are point sources at 1.6 \micron, which usually correspond to flat-spectrum sources.  This is additional evidence for a decrease in luminosity with evolution uncovered through a different means of classifying sources.

\subsubsection{The Stage 0 Lifetime\label{s.stage0}}

The distribution of protostars with respect to class can be used to estimate the lifetime of Class~0.  Assuming continuous star formation and a Class~II half-life of 2 Myr \citep{eva09}, \citet{dun14} estimated a combined protostellar (Class~0 and I) lifetime of 0.5 Myr.  \citet{dun15} accounted for additional sources of uncertainty in lifetime calculations. In particular, they argued that the Class II half-life of 2 Myr that is the basis for such calculations may be better estimated as 3 Myr, and that Class III objects that retain disks should be added to the Class II count. With these additional effects, plausible lifetimes for the protostellar phase extend from 0.46 to 0.72 Myr.  With 30\% of the HOPS protostars in Class~0, the implied Class~0 lifetime is 30\% of 0.5~Myr, or 0.15 Myr.  This extends from 0.14 to 0.22 Myr if the \citet{dun15} uncertainties are included.

As discussed in the Introduction, the SED class is not a perfect evolutionary indicator.  The physical {\em stage} of a YSO describes its actual evolutionary condition, which is only suggested by the observed {\em class} \citep{rob06,dun14}.  Although it is difficult to determine, it is worthwhile to investigate the lifetime of Stage 0, when the envelope mass is greater than the mass of the central star.  A Stage~0 lifetime that is short relative to the envelope lifetime would point to an early period of rapid mass accretion for protostars, suggesting that Stage I is a relatively long period of lower-level accretion punctuated by episodic bursts.  In this case, the true Stage 0 population might feature some very luminous but heavily extinguished young protostars.

In each model plotted in Figure~\ref{f.model}, we can determine when the central star reaches half its final mass, corresponding to the transition from Stage 0 to I.  For the models shown, these are 0.075, 0.070, and 0.063~Myr for the low-, \mbox{medium-}, and high-mass models, respectively.  These are less than the Class~0 duration of 0.14 to 0.22 Myr.  If we instead choose models in which half the stellar mass is assembled in about 0.15~Myr, matching the observationally derived Class~0 duration and requiring larger $t_H$, then the envelope masses inside 2500 AU are still of order $10^{-2}$ $M_\sun$ at 0.5~Myr, and many of the HOPS protostars correspond to models at times near 1 Myr, which is inconsistent with published estimates of the envelope lifetime.  This roughly 0.07~Myr Stage~0 lifetime is only about a factor of three greater than the 0.025~Myr lifetime estimated for the PBRS by \citet{stu13} based on the fraction of protostars in that category, assuming PBRS represent a distinct phase of star formation.

If the Stage 0 lifetime is shorter than the Class 0 lifetime, then some of the objects that are young according to observational diagnostics are really more evolved; i.e., some of the Class 0 sources are actually Stage I sources viewed at inclinations near edge-on.  The time it takes to assemble half the star is then shorter than estimated from SED analysis; i.e., several times $10^4$~yr instead of more than $10^5$ yr.  In this case, most of the envelope infall period of star formation would be characterized by a state of slowly declining, low-level accretion.

\subsection{Scatter in Luminosities: Episodic Accretion}

Exponentially declining infall rates that form a reasonable distribution of stellar masses can explain most of the scatter in the BLT and TLM diagrams. While episodic accretion is not a predominant factor in this scenario, it clearly occurs and is likely responsible for some of the spread.  The luminosity changes of V2775 Ori (HOPS 223), HOPS 383, and V1647 Ori (HOPS 388) were factors of 9.3, 35, and 9.7 \citep{fis12,saf15,and04}, while the ratio of the third quartile total luminosity to the first quartile total luminosity in each mass bin ranges from 6.5 to 17.  (See the blue line for HOPS 223 in Figure~\ref{f.blt}.)

It has been argued from radiative-transfer and hydrodynamical modeling \citep{dun12,vor15} and from the detection of CO$_2$ ice features as evidence of past high temperature \citep{kim12} that protostellar accretion outbursts must be frequent. E. J. Safron et al.\ (in preparation) searched for direct evidence of these outbursts by comparing IR photometry of the HOPS protostars at two epochs.  They find statistical evidence that protostars undergo hundreds of low-amplitude ($\sim10\times$) bursts during their formation periods. These outbursts would lead to scatter away from model tracks that lack episodic accretion.

For a more complete investigation of the influence of episodic accretion on the TLM diagram, we generated luminosity histograms from exponentially declining model tracks.  For each of the envelope mass bins in Table~\ref{t.medlum} except the most massive one, we randomly chose 1000 stars from an initial mass function. We used a function $dN/dM \propto M^{-\alpha}$, where $\alpha=0$ for $M_*<0.07$ $M_\sun$ \citep{all05}, $\alpha=1.05$ for $0.07\ M_\sun <M_*<0.5\ M_\sun$ \citep{kro02}, and $\alpha=2.35$ for $M>0.5$ $M_\sun$ \citep{kro02}.\footnote{This mass function is implemented in an IDL routine \texttt{cnb\_imf.pro} by C. Beaumont, available at \url{http://www.ifa.hawaii.edu/users/beaumont/code/cnb_imf-code.html}.} We then calculated the tracks through TLM space needed to yield stars of these masses and checked the luminosity along each track at a randomly selected mass within the bin. The half-life is assumed to be 0.06 Myr; varying this by 0.01 Myr in either direction has no appreciable effect on the results.

The model and data-derived histograms have about the same widths in all bins: the third quartile luminosity is about a factor of 10 greater than the first quartile luminosity.  The median luminosities differ, however.  The SED-derived luminosities (Table~\ref{t.medlum}) are 15.1, 2.56, 2.01, and 1.74 $L_\sun$, while the modeled ones are 3.63, 0.79, 0.35, and 0.33 $L_\sun$.  The former values are factors of 3 to 6 greater than the latter ones, although they follow the same trend of sharply decreasing then leveling out as the envelope mass diminishes. 

Besides episodic accretion, two other factors could play an additional role in creating these discrepancies.  First, the mass and radius of the central source are poorly understood at early times.  If our assumed ratios of $M_*/R_*$ are too small, then the predicted accretion luminosities will be too small.  Second, the model distributions are influenced by low-luminosity protostars ($<0.1~L_\sun$) that are not detected in our observations.  We demonstrated earlier that any incompleteness is not dependent on SED class, but protostars of sufficiently low luminosity at all stages may be missed.

Identifying cases of episodic accretion in the absence of a historical outburst is non-trivial.  It would not be evident from SED fitting, which cannot cleanly distinguish outbursts from truly massive and luminous objects.  A promising avenue for determining the outburst history of an object is to look instead for unusually extended C$^{18}$O emission as a sign of past heating \citep{jor15}.  Additional signs of recent outbursts may include a small disk mass or radius due to depletion by rapid accretion onto the star or a large cavity opening angle due to clearing by enhanced mass loss.  The \citet{fur16} fit to the HOPS 383 outburst gives both a small disk (5~AU in radius) and a large cavity opening angle (45$^\circ$), consistent with this scenario.  With spectroscopy of the accretion region in the inner disk, accretion rates may be determined, providing direct evidence for episodic accretion \citep[e.g.,][]{fis12}, and spectra indicative of an optically thick inner disk also point to outburst conditions \citep{con10}.

Our data and approach for Orion, a high-mass cloud, can be contrasted with that of \citet{dun10}. In their explanation of the BLT distribution of YSOs in five nearby low-mass molecular clouds, they postulated constant infall rates with the variation in luminosities mainly due to episodic accretion.  We instead postulate declining infall rates with a reduced but nonzero role for episodic accretion.  Our finding recalls the theoretical work of \citet{off11}, where models with constant accretion times and larger infall rates for larger final masses produce a broad protostellar luminosity function.  Continued analysis of the occurrence rate, magnitudes, and decay times of protostellar luminosity outbursts as a function of both cloud mass and environment will be crucial for understanding the importance of episodic accretion in explaining protostellar evolutionary diagrams.

\subsection{Variation as a Function of Environment:\\Changing Star-Formation Rates?}

The histogram of $T_{\rm bol}$ for the HOPS sample is remarkably flat.  The model-derived histogram of envelope mass is also flat for masses below 0.3 $M_\sun$.  Assuming a constant star-formation rate over the lifetime of protostars, we used this to argue for exponentially decreasing envelope densities.  The HOPS program deliberately selected out YSOs thought to be of Class II, and the number of sources in our sample with $T_{\rm bol}$ consistent with Class II is highly incomplete.  Typically in Orion, Class II sources outnumber protostars by a factor of three \citep{meg16}.

Although the assumption of a constant star-formation rate may be valid for the full sample, the $T_{\rm bol}$ and $M_{\rm env}$ histograms of the different regions within the Orion complex suggest that the star-formation rate may not be constant within each region.  The larger fraction of Class~0 protostars in Orion~B, 41\% as opposed to 35\% in the ONC and 20\% in L~1641, is consistent with the finding of \citet{stu13} that the fraction of deeply embedded PBRS is largest in Orion~B.  The larger fraction of Class~0 protostars in the ONC than in L~1641 is consistent with the finding of \citet{stu15} that the fraction of Class~0 protostars is larger in the north of Orion~A than in the south; these authors found that this fraction is correlated with the column density distribution shape (N-PDF) variation across Orion A.  Combined with the statistically larger luminosities for Class~0 protostars, this indicates that the protostellar luminosity may be rooted in the molecular cloud N-PDF structure, which in turn is rooted in the density structure of the star-forming material (see, e.g., \citealt{stu16}). That is, regions with more local (or centrally concentrated and potentially filamentary) mass may produce higher-luminosity protostars.

These variations are evident in the $T_{\rm bol}$ and $M_{\rm env}$ histograms.  The Orion~B cloud shows a distinct decrease in the number of protostars with increasing evolutionary state,  progressing from the young Class~0 or Stage~0 protostars, with low $T_{\rm bol}$ and high $M_{\rm env}$, to more evolved Class~I or Stage~I protostars.  This decline may be the result of an increasing star-formation rate, with the rate increasing by roughly a factor of three over the protostellar lifetime.  Although the low star-formation efficiency of Orion B \citep{meg16} may suggest that it is younger than Orion A, observations of pre-main-sequence stars show that star formation has occurred over 2 Myr in this cloud \citep{fla08}.  Thus, the increase in the star-formation rate may be recent.  The higher fraction of Class~0 protostars in the ONC may also be the result of a recent increase in the star-formation rate.  In contrast, L 1641 shows a hint of a decrease in the star-formation rate.  In general, these diagrams suggest that although the star-formation rate over the last 0.5 Myr is relatively stable when considered over the entire cloud, it may not be as constant when considering smaller regions within it.

\section{CONCLUSIONS}

We have determined the bolometric luminosities and temperatures for 330 YSOs, 315 of which have $T_{\rm bol}$ consistent with protostars, in the Orion molecular clouds that were targets of HOPS, the \Herschel\ Orion Protostar Survey.  The $L_{\rm bol}$ histogram is broad, ranging over nearly five orders of magnitude, with a peak near 1 $L_\sun$, a width at half-maximum of two orders of magnitude, and a tail extending beyond 100 $L_\sun$.  The $T_{\rm bol}$ histogram is flat, with logarithmic bins between 30 and 600~K each containing approximately equal numbers of protostars.  The BLT diagram features a broad spread in luminosities at each bolometric temperature, with 29\% of the sources having $T_{\rm bol}<70$ K, the dividing line for Class~0 and Class~I protostars.

The BLT diagram shows a systematic decline in the median luminosity with increasing bolometric temperature.  This decline is reflected in different luminosities for the Class~0 and Class~I protostars.  The median Class~0 luminosity is 2.3~$L_\sun$ compared to 0.87~$L_\sun$ for Class~I, indicating that more deeply embedded protostars are more luminous. The Class~0 luminosity histogram has less than a 0.1\% probability of being drawn from the same underlying distribution as the Class~I luminosity histogram.

We divided the sample into regions; from south to north, these are L~1641, the ONC, and Orion~B.  The fraction of protostars in Class~0 increases from south to north, from 20\% in L~1641 to 41\% in Orion B.  Within each region, trends seen in the entire sample persist.  The Class 0 protostars are statistically brighter than the Class I protostars, and $L_{\rm bol}$ declines with $T_{\rm bol}$ while having a large dispersion.  We argued that these trends are unlikely to be driven by incompleteness or an inaccurate accounting for foreground extinction.  In the ONC and Orion B, there is a decrease in the number of protostars at progressively larger $T_{\rm bol}$, suggesting a relatively recent increase in the star-formation rate.

These findings are further confirmed via our SED-fitting analysis.  When we fit SED models to the protostars and Class~II objects to estimate their total luminosities and envelope masses inside 2500 AU, the median total luminosity increases to 3.0~$L_\sun$.  The mass inside 2500~AU is not the entire mass of the envelope; it is only the portion that is warm enough to be traced by our far-IR measurements.  In all but the earliest stages, this gas is falling toward the central protostar and is continually replenished by gas from outside 2500~AU.

Trends in total luminosity and envelope mass are similar to those in $L_{\rm bol}$ and $T_{\rm bol}$. The histogram of envelope masses is quite flat, luminosities are largest for envelope masses between 0.1 and 1 $M_\sun$ and fall as the envelopes become less massive, and there is a spread in luminosity of about three orders of magnitude in each mass bin.  The flat histogram of envelope mass and the decrease in luminosity can largely be explained by an overall exponential decrease in the envelope infall rate with time as postulated by \citet{bon96}, \citet{mye98}, \citet{sch04}, and \citet{vor10}.  We show that simple models invoking an exponentially decreasing envelope mass can approximately reproduce most aspects of the observed distribution of sources in an $L_{\rm tot}$ versus $M_{\rm env}$ diagram.  In these models, we find that the time to assemble half the star, which corresponds to the time of the physical transition from Stage 0 to Stage I, is about half the observationally derived Class~0 lifetime.

The initial mass of the envelope and the half-life of the envelope mass set the final mass of the star.  When the exponentially declining models are applied to an ensemble of cases that yield a typical initial mass function of main-sequence stars, the luminosities in each mass bin have a similar spread to those derived from the data but are systematically lower.  In this model, the distribution of luminosities at each envelope mass is largely due to the expected distribution in the final masses of the forming stars.

\acknowledgments

Support for this work was provided by the National Aeronautics and Space Administration (NASA) through awards issued by the Jet Propulsion Laboratory/California Institute of Technology (JPL/Caltech). This work is based on observations made with the \Spitzer\ Space Telescope, which is operated by JPL/Caltech under a contract with NASA; it is also based on observations made with the \Herschel\ Space Observatory, a European Space Agency Cornerstone Mission with significant participation by NASA. We include data from the Atacama Pathfinder Experiment, a collaboration between the Max-Planck Institut f\"ur Radioastronomie, the European Southern Observatory, and the Onsala Space Observatory. Finally, this publication makes use of data products from the Two Micron All Sky Survey, which is a joint project of the University of Massachusetts and the Infrared Processing and Analysis Center/Caltech, funded by NASA and the National Science Foundation.

This work was supported by NASA Origins of Solar Systems grant 13-OSS13-0094. The work of WJF was supported in part by an appointment to the NASA Postdoctoral Program at Goddard Space Flight Center, administered by the Universities Space Research Association through a contract with NASA. JJT acknowledges past support from grant 639.041.439 from the Netherlands Organisation for Scientific Research (NWO) and from NASA through Hubble Fellowship grant \#HST-HF-51300.01-A awarded by the Space Telescope Science Institute, which is operated by the Association of Universities for Research in Astronomy, Inc., for NASA, under contract NAS 5-26555. AS is thankful for funding from the ``Concurso Proyectos Internacionales de Investigaci\'on, Convocatoria 2015'' (project code PII20150171) and the BASAL Centro de Astrof\'isica y Tecnolog\'ias Afines (CATA) PFB-06/2007.  MO acknowledges support from MINECO (Spain) grant AYA2014-57369-C3 (co-funded wth FEDER funds).

\end{document}